\documentclass[acmsmall,nonacm]{acmart}
\usepackage{subcaption}
\usepackage{multirow}
\usepackage{amsmath}
\usepackage{xcolor}
\usepackage{colortbl}
\usepackage{adjustbox}
\usepackage{fancyhdr}
\usepackage{setspace}
\usepackage{array,multirow}
\usepackage{makecell, tabularx}
\usepackage{color,soul}
\usepackage[most]{tcolorbox}
\usepackage{rotating, graphicx}
\usepackage{titlesec}
\usepackage{enumitem}
\usepackage{algorithmic}
\usepackage{graphicx}
\usepackage{textcomp}
\usepackage{xcolor}
\usepackage{comment}
\usepackage{titlesec}
\usepackage{stfloats}
\usepackage{enumitem}
\usepackage{booktabs} 
\usepackage{caption}
\usepackage{multirow} 
\usepackage{float}
\usepackage{color}
\usepackage{setspace}
\usepackage{tikz}

\setlist[itemize]{noitemsep, topsep=0pt}
\setlist[enumerate]{noitemsep, topsep=0pt}

\tcbset{
    left=8pt,
    right=8pt,
    enhanced,
    halign=center,
    colback=gray!16,
    colframe=black,
    boxrule=0.8pt,
    width=\dimexpr\textwidth\relax,
}

\usepackage{color}
\definecolor{darkred}{rgb}{0.55, 0.0, 0.0}
\definecolor{codegray}{rgb}{0.5,0.5,0.5}
\definecolor{mygreen}{RGB}{68,85,37}

\AtBeginDocument{%
  \providecommand\BibTeX{{%
    \normalfont B\kern-0.5em{\scshape i\kern-0.25em b}\kern-0.8em\TeX}}}

\setcopyright{acmcopyright}
\copyrightyear{2023}
\acmYear{2023}
\acmDOI{}

\acmJournal{POMACS}
\acmVolume{7}
\acmNumber{1}
\acmArticle{}
\acmMonth{3}

\newcommand{\disaggrSystems}{disaggregated systems}

\newcommand{\disaggrSystem}{disaggregated system}
\newcommand{\computeComponent}{compute component}
\newcommand{\computeComponents}{compute components}
\newcommand{\memoryComponent}{memory component}
\newcommand{\memoryComponents}{memory components}
\newcommand{\localMemory}{local memory}
\newcommand{\remoteMemory}{remote memory}

\newcommand{\PQ}{PQ}
\newcommand{\LC}{LC}
\newcommand{\BP}{BP}

\newcommand{\LocalOnly}{Local}
\newcommand{\RemoteOnly}{Remote}
\newcommand{\netlat}{\revised{switch latency}}
\newcommand{\netlats}{\revised{switch latencies}}
\newcommand{\bwf}{bandwidth factor}
\newcommand{\bwfs}{bandwidth factors}
\newcommand{\myName}{\emph{DaeMon}}

\newcommand*\circled[1]{\tikz[baseline=(char.base)]{\node[shape=circle,fill,inner sep=1.4pt] (char) {\textbf{\textcolor{white}{#1}}};}}

\setlist[itemize]{noitemsep, topsep=0pt}

\newcommand\cgiannou[1]{\noindent{\color{teal}}} 

\definecolor{amber}{rgb}{1.0, 0.49, 0.0}
\definecolor{darkgreen}{rgb}{0.0, 0.2, 0.13}
\definecolor{darkbyzantium}{rgb}{0.36, 0.22, 0.33}
\definecolor{darkseagreen}{rgb}{0.56, 0.74, 0.56}
\definecolor{darkspringgreen}{rgb}{0.09, 0.5, 0.27}
\definecolor{dollarbill}{rgb}{0.52, 0.73, 0.4}
\definecolor{MidnightBlue}{RGB}{46,46,208}

\newcommand\shepherd[1]{\noindent{\color{black}{#1}}} 
\newcommand\revised[1]{\noindent{\color{black}{#1}}} 
\newcommand\rebuttal[1]{\noindent{\color{black}{#1}}} 
\newcommand\fixed[1]{\noindent{\color{black}{#1}}} 
\begin{document}

\title{\myName: Architectural Support for Efficient Data Movement in Disaggregated Systems}

\author{Christina Giannoula}
\email{christina.giann@gmail.com}
\affiliation{%
  \institution{University of Toronto}
  \country{Canada}
}  
\affiliation{%
  \institution{National Technical University of Athens}
  \country{Greece}
}

\author{Kailong Huang}
\affiliation{%
  \institution{University of Toronto}
  \country{Canada}
}
\authornote{Equal contribution to this work.}

\author{Jonathan Tang}
\affiliation{%
  \institution{University of Toronto}
  \country{Canada}
}
\authornotemark[1]

\author{Nectarios Koziris}
\affiliation{%
  \institution{National Technical University of Athens}
  \country{Greece}
}

\author{Georgios Goumas}
\affiliation{%
  \institution{National Technical University of Athens}
  \country{Greece}
}

\author{Zeshan Chishti}
\affiliation{%
  \institution{Intel Corporation}
  \country{USA}
}

\author{Nandita Vijaykumar}
\affiliation{%
  \institution{University of Toronto}
  \country{Canada}
}

\renewcommand{\shortauthors}{Christina Giannoula, et al.}

\begin{abstract}
%\vspace{-24pt}
Resource disaggregation offers a cost effective solution to resource scaling, utilization, and failure-handling in data centers by physically separating hardware devices in a server. Servers are architected as pools of \shepherd{processor}, memory, and storage devices, organized as independent failure-isolated components interconnected by a high-bandwidth network. A critical challenge, however, 
is the high performance penalty of accessing data from a remote memory module over the network. Addressing this challenge is difficult as disaggregated systems have high runtime variability in network latencies/bandwidth, \shepherd{and page migration can significantly delay critical path cache line accesses in other pages.}

%Addressing this challenge is difficult as disaggregated %architectures
%\revised{systems} have high runtime variability in network latencies/bandwidth. To alleviate data access costs, existing proposals migrate pages to a small local memory in the compute unit during a remote memory access. Page migration however incurs significant performance overheads by slowing down critical path accesses to cache lines in other pages. 

This paper \rebuttal{conducts a characterization analysis on different data movement strategies in fully disaggregated systems, evaluates their performance overheads in a variety of workloads, and } introduces \myName{}, the first software-transparent mechanism to \rebuttal{significantly} alleviate data movement overheads in fully disaggregated systems. First, \revised{to enable scalability to multiple hardware components in the system, we enhance each compute and memory unit with specialized engines that transparently handle data migrations.} Second, \fixed{to achieve high performance and provide robustness across various network, architecture and application characteristics, we implement a synergistic approach of bandwidth partitioning, link compression, decoupled data movement of multiple granularities, \shepherd{and} adaptive granularity selection in data movements.}  We evaluate \myName{} \shepherd{in a wide variety of workloads at different network and architecture configurations using a state-of-the-art simulator}. \myName{} improves system performance and data access costs by 2.39$\times$ and 3.06$\times$, respectively, over the \shepherd{widely-adopted approach of moving data at page granularity.}

%This paper introduces \myName{}, the first software-transparent mechanism to alleviate data movement overheads in fully disaggregated systems. First, \revised{to enable scalability to multiple hardware components in the system, we enhance each compute and memory unit with specialized engines that transparently handle data migrations.} Second, \fixed{to achieve high performance and provide robustness across various network, architecture and application characteristics, we implement a synergistic approach of bandwidth partitioning, link compression, decoupled data movement of multiple granularities, adaptive granularity selection in data movements.}  We evaluate \myName{} using a wide variety of workloads at different network and architecture configurations. \myName{} improves system performance and data access costs by 2.39$\times$ and 3.06$\times$, respectively, over the baseline system.

\vspace{-2pt}
\end{abstract}


\keywords{data movement, data access, memory access, hardware support, hardware mechanism, high performance, memory systems, memory disaggregation, resource disaggregation, disaggregated systems, workload characterization, benchmarking, performance characterization}

\maketitle

\section{Introduction}

\begin{figure}[t]
    %\vspace{-2pt}
    \centering
    \includegraphics[width=1\linewidth]{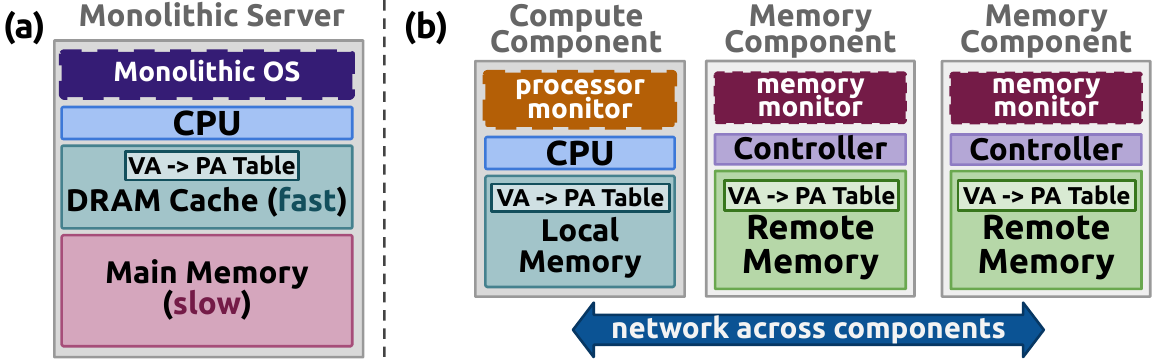}
    \vspace{-10pt}
    \caption{(a) A monolithic system versus (b) a disaggregated system.}
    \label{fig:baseline-architectures-diff}
    \vspace{-10pt}
\end{figure}

With recent advances in network technologies~\cite{rdma_melanox,rdma,omni_path,genz,Gao2016Network} that enable high bandwidth networks, \textit{resource disaggregation}~\cite{Shan2018LegoOS,Gao2016Network} has emerged as a promising technology for data centers~\cite{Shan2018LegoOS,Lee2021MIND,Calciu2021Rethinking,Gao2016Network,Tsai2017,guo2021clio,Giannoula2022Thesis}. Resource disaggregation proposes the physical separation of hardware devices (CPU, accelerator, memory, and disk) in a server as independent and failure-isolated components connected over a high-bandwidth network such as RDMA~\cite{rdma_melanox} and Gen-Z~\cite{genz}. Compared to monolithic servers that tightly integrate these components (Figure~\ref{fig:baseline-architectures-diff}a), \emph{\textbf{\disaggrSystems{}}} %disaggregated systems (\textbf{\disaggrSystems{}}) 
can greatly improve \emph{resource utilization}, as memory/storage components can be shared across applications; \emph{resource scaling}, as hardware components can be flexibly added, removed, or upgraded; and \emph{failure handling}, as the entire server does not need to be replaced in the event of a fault in a device. Thus, resource disaggregation can significantly decrease data center costs.

Disaggregated systems comprise multiple compute, memory and storage components, interconnected over a high-bandwidth network  (Figure~\ref{fig:baseline-architectures-diff}b), each independently managed by a specialized kernel module (monitor). Typically, each \emph{\textbf{\computeComponent{}}} %compute component (\textbf{\computeComponent{}}) 
includes a small amount (a few GBs) of main memory (henceforth referred to as \emph{\textbf{\localMemory{}}}) %\emph{local memory} (\textbf{\localMemory{}})) 
to improve memory performance. \revised{However, almost all the memory in the data center is separated as network-attached \textit{disaggregated memory} components to maximize resource sharing and independence in failure handling (different from typical hybrid memory architectures). Thus, the \emph{majority} of the application working sets \shepherd{is} accessed from the \textit{disaggregated memory} components (henceforth referred to as \textbf{\emph{remote memory}}}). %\emph{remote memory} (\textbf{\remoteMemory{}})).} 
Each \emph{\textbf{\memoryComponent{}}} %memory component (\textbf{\memoryComponent{}}) 
includes its own controller and can be flexibly shared by many \computeComponents{}. Thus, \disaggrSystems{} can provide high memory capacity for applications with large working sets (e.g., bioinformatics, graph processing and neural networks) at lower cost. Fine-grain microsecond-latency networking technologies~\cite{rdma_melanox,rdma,omni_path,genz,Chuanxiong2016RDMA} that interconnect all hardware components have made fully disaggregated systems feasible, being only 2-8$\times$ slower than DRAM bus bandwidth. However, since a large fraction of the application's data (typically $\sim$80\%)~\cite{Shan2018LegoOS,Lee2021MIND,Gao2016Network} is located and accessed from \remoteMemory{}, the higher latencies of remotely accessing data over the network can cause large performance penalties.

Alleviating data access overheads is challenging in \disaggrSystems{} for the following reasons. First, \disaggrSystems{} are not monolithic and comprise independently managed entities: each component has its own hardware controller, its resource allocation is transparent from other components and a specialized kernel monitor uses its own functionality/implementation to manage the component it runs on (only communicates with other monitors via network messaging if there is a need to access remote resources). This characteristic necessitates a distributed and disaggregated solution %for data movement 
that can scale to a large number of independent components in the system. Second, there is high variability in data access latencies as they depend on the location of the remote \memoryComponent{} and the contention with other \computeComponents{} that share the same \memoryComponents{} and network. Data placements can also vary during runtime or between multiple executions, since data is dynamically allocated in one or more remote \memoryComponents{} and hardware updates can flexibly change the architecture of the \memoryComponent{} and the network topology. Third, data is typically migrated at page granularity \cite{Yan2019Nimble,Aguilera2017Remote,Zhang2020RethinkingDM,Lim2012System,Shan2018LegoOS,Angel2020Disaggregation,Gu2017Infiniswap,Aguilera2018Remote,Lee2021MIND} as it enables: (i) transparency to avoid modifications to existing OS/applications; (ii) low metadata overheads for address translation; and (iii) leveraging spatial locality within pages. However, we observe in §\ref{MotivationPlotbl} that moving memory pages in \disaggrSystems{}, i.e., moving data at a large granularity over the network, can significantly increase bandwidth consumption and slow down accesses to cache lines in other concurrently accessed pages.

Recent works on hybrid memory systems~\cite{Dong2010Simple,Doudali2019Kleio,Yan2019Nimble,Liu2017Hardware,Kotra2018Chameleon,Agarwal2017Thermostat,Mitesh2015Heterogeneous,Ruan2020AIFM,Dulloor2016Data,Agarwal2015Page,Kim2021Exploring,Doudali2021Cori,Lei2019Hierarchical,Sudarsun2017HeteroOS,Loh2011Efficiently,Loh2012Supporting,Chou2014Cameo,Ryoo2017Silcfm,Jevdjic2013Footprint}, for example, those that integrate die-stacked DRAM~\cite{HBM} caches aim to address the high page movement costs between main memory and the DRAM cache~\cite{Loh2011Efficiently,Loh2012Supporting,Chou2014Cameo,Ryoo2017Silcfm,Jevdjic2013Footprint} with mechanisms to move data at smaller granularities~\cite{Vasilakis2019LLCGuided,Jiang2010CHOP,Prodromou2017MemPod,Loh2011Efficiently,Loh2012Supporting,Chou2014Cameo,Vasilakis2020Hybrid2}, e.g., cache line, or by using page placement/hot page selection mechanisms~\cite{Dong2010Simple,Doudali2019Kleio,Yan2019Nimble,Liu2017Hardware,Kotra2018Chameleon,Agarwal2017Thermostat,Mitesh2015Heterogeneous,Ruan2020AIFM,Dulloor2016Data,Agarwal2015Page,Kim2021Exploring,Doudali2021Cori,Lei2019Hierarchical,Sudarsun2017HeteroOS}. However, these prior works are tailored for a monolithic tightly-integrated architecture (Figure~\ref{fig:baseline-architectures-diff}a), and are not suitable for \disaggrSystems{} (See §~\ref{RelatedWorkbl}). These works assume \emph{centralized} data management/allocation (unlike in \disaggrSystems{}). %Most of them assume that the memory management/ and%resource 
%allocation at \emph{different} memory tiers are \emph{centralized} and fully controlled by the CPU side. 
For instance, software runtimes~\cite{Dong2010Simple,Doudali2019Kleio,Yan2019Nimble,Liu2017Hardware,Kotra2018Chameleon,Agarwal2017Thermostat,Mitesh2015Heterogeneous,Ruan2020AIFM,Dulloor2016Data,Agarwal2015Page,Kim2021Exploring,Doudali2021Cori,Lei2019Hierarchical,Sudarsun2017HeteroOS} running on CPUs in hybrid systems leverage TLBs/page tables to track \fixed{page} hotness \fixed{and move} pages %placed 
across different memory devices (Figure~\ref{fig:baseline-architectures-diff}a). Instead, in fully \disaggrSystems{} all hardware memory functionalities (e.g., TLBs, page tables) of remote pages are moved to the \memoryComponents{} themselves~\cite{Shan2018LegoOS, guo2021clio} (Figure~\ref{fig:baseline-architectures-diff}b). Thus they cannot be used to track page hotness at the CPU \shepherd{side} to implement intelligent page placement/movement in \localMemory{}. Similarly, hardware-based approaches~\cite{Vasilakis2019LLCGuided,Jiang2010CHOP,Prodromou2017MemPod,Li2017Utility} add \emph{centralized} hardware units in the CPU to track metadata for pages in second-tier memory. This however would incur high hardware costs in \disaggrSystems{} that enable large amounts (e.g., TBs) of \remoteMemory{}~\cite{Shan2018LegoOS}. Requiring \emph{each} \computeComponent{} to control/track a large number of pages in remote \memoryComponents{} would impose significant hardware costs and scalability challenges, and thus \fixed{might} annihilate the benefits of resource disaggregation. %, thus necessitating a distributed approach.
%Furthermore, these prior works do not handle variability in data access costs.
\fixed{Moreover,} \disaggrSystems{} incur significant variations in access latencies and bandwidth based on the current network architecture and concurrent jobs sharing the \memoryComponents{}/network, \shepherd{which are} not addressed by prior work. This necessitates a solution primarily designed for robustness to this variability.

In this work, we \rebuttal{analyze different data movement strategies in fully \disaggrSystems{}, and} introduce \myName{}, \rebuttal{an efficient} \shepherd{software-transparent} mechanism to alleviate data movement overheads in \disaggrSystems{}. \myName{} provides (i) high performance on dynamic workload demands, (ii) robustness to variations in architectures, network characteristics and application behavior, and (iii) independence and scalability to multiple \computeComponents{} and \memoryComponents{} that are managed transparently to each other and are flexibly added/removed in the system. %\cgiannou{DaeMon is designed to achieve the goals of performance, software-transparency robustness to dynamic variations in architectures, network characteristics and scalability to multiple to multiple \computeComponents{} and \memoryComponents{}.}

%\myName{} is designed to achieve the goals of performance, low-cost page migration, software-transparency and robustness through \revised{four} key \revised{design options}. 

\myName{} \fixed{consists of two key ideas.} First, we offload data migrations to dedicated hardware engines, named \myName{} compute and memory engine, that are added at each \computeComponent{} and \memoryComponent{}, respectively. This key idea enables independence
%, \fixed{in keeping with the promise of resource disaggregation,} 
and scalability to a large number of \computeComponents{} and \memoryComponents{} of \disaggrSystems{}. \fixed{Compared to a centralized design, \myName{}'s distributed management of data movement enables simultaneous processing of data movement across multiple components and decreases the processing costs and queuing delays to serve data requests.} Second, \fixed{we leverage the synergy of three key techniques to provide robustness to the high variability in network latencies/bandwidth.} 1) We use a bandwidth partitioning approach to enable the decoupled movement of data at \emph{two granularities}, i.e., page and cache line, and \emph{prioritize} cache line granularity data moves over page moves. This design enables low access latencies to \remoteMemory{} for the cache line requests on the critical path, while the associated pages can be still be moved independently at slower rates to retain the benefits of spatial locality. 2) We design an adaptive approach to decide on-the-fly if a request should be served by a cache line, page or \emph{both}. Via selective granularity data movement, we provide robustness to variations in network, architectures and application characteristics. 3) We leverage hardware link compression when migrating pages to reduce network bandwidth consumption and alleviate queuing delays.

The \fixed{synergy of the aforementioned key techniques provides a robust solution for \disaggrSystems{}:}
decoupled multiple granularity data movement effectively prioritizes cache line requests on the critical path, and migrates pages at a slower rate leveraging compression to reduce bandwidth consumption. The adaptive granularity selection mechanism effectively adapts to \shepherd{the characteristics of the} application data, e.g., by favoring moving more pages if \shepherd{application} data is highly compressible. The decoupled cache line granularity movement also enables the use of more sophisticated and effective compression algorithms (\shepherd{with relatively} high compression latency) for page migrations.

We evaluate \myName{} using a range of capacity intensive workloads with different memory access patterns from machine learning, high-performance-computing, graph processing,  and bioinformatics domains.  Over the widely-adopted approach of moving data at page granularity, \myName{} decreases memory access latencies by 3.06$\times$ on average, and improves system performance by 2.39$\times$ on average. We \shepherd{demonstrate} that \myName{} provides (i) robustness and significant performance benefits on various network/architecture configurations and application behavior (Figures~\ref{fig:single-performance} and ~\ref{fig:disturbance}), (ii) scalability to \emph{multiple} hardware components and networks, %improving the system performance by 3.25$\times$ 
(Figure~\ref{fig:multimem}), and (iii) adaptivity to dynamic workload demands, even when multiple heterogeneous jobs are concurrently executed in the \disaggrSystem{} (Figure~\ref{fig:multibench}).

%We demonstrate that \myName{}'s synergistic techniques provide robustness and significant performance benefits on various network/architecture configurations and application behavior (§\ref{Evaluationbl}), and even when multiple jobs are concurrently executed in the \disaggrSystem{} (Figure~\ref{fig:multibench}).

This paper makes the following contributions:
\begin{itemize}[noitemsep,topsep=0pt,leftmargin=8pt]
    \item \rebuttal{We heavily modify a \shepherd{state-of-the-art} simulator to develop and evaluate the overheads of different data movement strategies in fully \disaggrSystems{}, analyze the challenges of providing efficient data movement in such systems, and develop \myName{},  an adaptive distributed data movement mechanism for fully \disaggrSystems{}.}
    %\item We investigate the challenges of providing efficient data movement in fully \disaggrSystems{}, and propose an adaptive distributed mechanism, \myName{}, for such systems, that provides robustness under varying network/application characteristics.
    \item We enable decoupled data movement at two granularities, and migrate the requested critical data quickly at cache line granularity and the corresponding pages opportunistically without stalling critical cache line \shepherd{requests}. We dynamically control the data movement granularity to effectively adapt to the current system load and application behavior. We employ a high-latency compression scheme to further reduce bandwidth consumption during page migrations.
    \item We evaluate \myName{} using a wide range of capacity intensive workloads, various architecture/network configurations, and in multi-workload executions of concurrent heterogeneous jobs. We demonstrate that \myName{} significantly outperforms the state-of-the-art \shepherd{data movement strategy,} and constitutes a robust and scalable approach \fixed{for data movement in \shepherd{fully} \disaggrSystems{}.} %across various network/architecture/application characteristics.
\end{itemize}
%\vspace{-6pt}
\section{Background and Motivation}
%\vspace{-5pt}
\subsection{Baseline \shepherd{Disaggregated} System}
%\vspace{-1pt}

\begin{figure}[t]
    %\vspace{-8pt}
    \centering
    \includegraphics[width=1\linewidth]{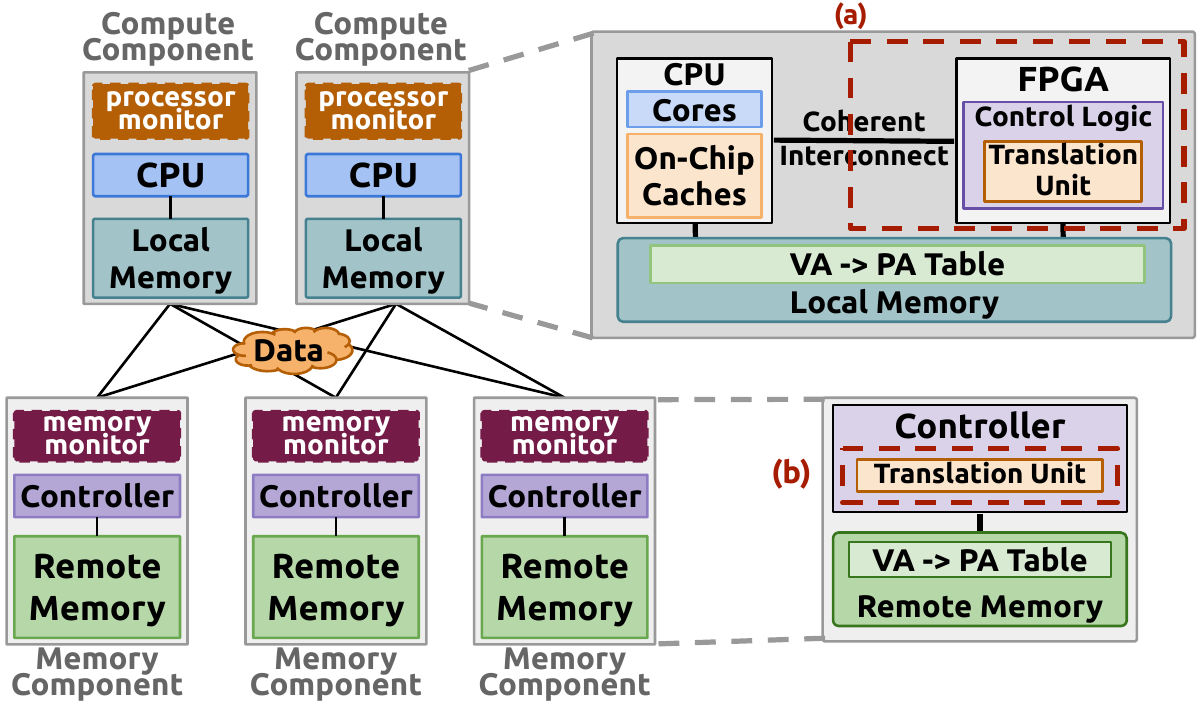}
    \vspace{-14pt}
    \caption{High-level organization of a \disaggrSystem.}
    \label{fig:baseline-architecture}
    \vspace{-6pt}
\end{figure}

Figure~\ref{fig:baseline-architecture}
shows the baseline organization of the \disaggrSystem{}, which includes several \computeComponents{} and \memoryComponents{} as network-attached components. To improve performance, each \computeComponent{} tightly includes a few GBs of main memory, referred to as \textbf{\localMemory{}}, %\emph{local memory} (\textbf{\localMemory{}}), 
which can typically host 20-25\% of the application's memory 
footprint~\cite{Shan2018LegoOS,Gao2016Network,sapio2017daiet}. Each \memoryComponent{} includes its own controller and connects multiple DIMM modules, referred to as \textbf{\remoteMemory}. %\emph{remote memory} (\textbf{\remoteMemory}).

We assume distributed OS modules \shepherd{that coordinate and communicate with each other via network messaging when needed}, similar to ~\cite{Shan2018LegoOS,Lee2021MIND,guo2021clio}: processor and memory kernel monitors run at \computeComponents{} and \memoryComponents{}, respectively. The memory allocation/management of \remoteMemory{} is performed at the \memoryComponent{} \shepherd{itself}~\cite{guo2021clio,Shan2018LegoOS}, transparently to \computeComponents{}, enabling the different components to be \emph{independent}. The on-chip caches and the \localMemory{} of \computeComponents{} are typically indexed by virtual addresses~\cite{Shan2018LegoOS}, and remote data is requested from \memoryComponents{} using virtual addresses~\cite{Shan2018LegoOS,Lee2021MIND,Calciu2021Rethinking,guo2021clio} \shepherd{(unlike in hybrid memory systems)}. %\cgiannou{(Note that \myName{} could also work with physically-indexed caching schemes.)} 
The data management is typically performed at page granularity~\cite{Calciu2021Rethinking,Shan2018LegoOS,Lee2021MIND} (e.g., 4KB). The \localMemory{} of \shepherd{the} \computeComponent{} can be treated as a cache with tags~\cite{Shan2018LegoOS} or a \emph{local} virtual to physical translation mapping~\cite{Lee2021MIND} can be used (either approach works with \myName{}, however we assume the second approach in our evaluation). The physical memory \shepherd{addresses} of \shepherd{the} \localMemory{} \shepherd{can be found by accessing and traversing} metadata (tags or local page tables) kept in a dedicated (pre-reserved) DRAM memory \shepherd{space} (kernel metadata is directly indexed via physical addresses). 
%The physical memory of \localMemory{} is accessed using metadata (tags or local page tables) kept in dedicated (pre-reserved) DRAM memory (kernel metadata is directly indexed via physical addresses). 
When an \localMemory{} \emph{miss} happens, either (i) the processor kernel module of the \computeComponent{} triggers a page fault and fetches the requested page from remote \memoryComponents{}~\cite{Shan2018LegoOS}, or (ii) dedicated software runtimes co-designed with hardware primitives~\cite{Calciu2021Rethinking}  (e.g., supported in FPGA-based controllers as shown in Figure~\ref{fig:baseline-architecture}a) handle remote data requests on demand completely eliminating expensive page faults. Either approach works with \myName{}. 
\revised{We} assume that \revised{the controllers of \memoryComponents{} implement} \emph{hardware-based} address translation (Figure~\ref{fig:baseline-architecture}b) %can be supported 
to access \revised{pages in} \remoteMemory{} %of \memoryComponents{} 
\revised{as proposed in}~\cite{guo2021clio}.

Jobs running at different \computeComponents{} can share \emph{read-only} pages located at multiple \memoryComponents{}. \shepherd{Similarly to} prior state-of-the-art 
works~\cite{Shan2018LegoOS,guo2021clio,Calciu2021Rethinking}, we assume that the system does not support writable shared pages across \computeComponents{}, since they are rare across datacenter \revised{jobs}~\cite{Shan2018LegoOS,guo2021clio}.

%\vspace{-5pt}
%\subsection{Performance Overheads in Data Movement}\label{MotivationPlotbl}
\subsection{\shepherd{Data Movement Overheads in Fully Disaggregated Systems}}\label{MotivationPlotbl}
%\vspace{-1pt}

Prior state-of-the-art works~\cite{Chenxi2020Semeru,Pengfei2021OneSided,Zhang2020RethinkingDM,Bindschaedler2020Hailstorm,Peng2020Underutilization,Pinto2020ThymesisFlow,Katrinis2016dredbox,guo2021clio,Gu2017Infiniswap,Lee2021MIND} typically enable data management at page granularity for three compelling reasons. First, the memory allocation and management is \textit{transparent}, i.e., requires little to no modification to OS or application code. Second, the coarse granularity enables \textit{low metadata overheads} for address translation in \localMemory{} and \remoteMemory. Managing \localMemory{} as a cache at cache line granularity would incur prohibitively high metadata overheads~\cite{Shan2018LegoOS}. Third, page \shepherd{movements} enable exploiting \emph{spatial locality} in common memory access patterns~\cite{Lim2009Disaggregated,Jiang2010CHOP,Woo2010AnOptimized}, and increase the number of accesses served from the lower cost \localMemory{} instead of \remoteMemory.

Figure~\ref{fig:motivational-results} compares the performance of different data movement \shepherd{strategies} in \disaggrSystems{} \shepherd{across various workloads (See Table~\ref{table:workloads})}. We evaluate one \memoryComponent{} and one \computeComponent{} having \localMemory{} to fit $\sim$20\% of the application's \shepherd{working set}. We use 100 ns/400 ns latency~\cite{Gao2016Network,Lee2021MIND} to model the propagation and switching delays on the network (referred to as \textbf{\emph{\netlat}}), and configure the network bandwidth \shepherd{between the \computeComponent{} and the \memoryComponent{} (referred to as \textbf{\emph{\bwf}})} to be 1/4$\times$ the DRAM \shepherd{bus} bandwidth~\cite{Shan2018LegoOS,Gao2016Network} \shepherd{of the \localMemory{} or \remoteMemory}. We compare six configurations: (i) \LocalOnly: all accesses are served from \shepherd{the} \localMemory{}; (ii) \emph{cache-line}: accessing data from \remoteMemory{} at cache line granularity, and directly writing data to \shepherd{the Last Level Cache (LLC) of the \computeComponent{}} (\localMemory{} is not used), (iii) \RemoteOnly: accessing data from \remoteMemory{} at page granularity (moving pages to \localMemory{}) \shepherd{accounting for} all network-related overheads, (iv) \emph{page-free}: remote accesses incur the latency of \shepherd{one cache line granularity remote access} %a remote cache line access 
and the corresponding page is transferred to \localMemory{} at \emph{zero cost} (spatial locality \revised{is} leveraged), (v) \emph{cache-line+page}: requesting data from \remoteMemory{} at both cache line (moved to LLC) and page granularity (moved to \localMemory{}) and servicing data requests using the latency of the packet that arrives earlier to \computeComponent{} (\shepherd{accounting for} all network-related overheads), and (vi) \emph{\myName{}}: accessing data from \remoteMemory{} using \myName{}  (\shepherd{accounting for} all network-related overheads).

\begin{figure}[H]
    %\vspace{-1pt}
    \centering
    \includegraphics[width=\linewidth]{results/motivation_overview_netlat_100_bwsf_4_ipc_slowdown.pdf}
    \centering
    \includegraphics[width=\linewidth]{results/motivation_overview_netlat_400_bwsf_4_ipc_slowdown.pdf}
    \vspace{-16pt}
    \caption{Data movement overheads in \disaggrSystems{}.}
    \label{fig:motivational-results}
    \vspace{-6pt}
\end{figure}

We make four observations. First, \RemoteOnly{}, i.e., the typically-used approach of moving data at page granularity, incurs significant performance slowdowns, on average 3.86$\times$, over the monolithic \LocalOnly{} configuration due to transferring large amounts of data over the network. In addition to the large network bandwidth consumption, migrating pages can slow down critical path accesses to data in other \shepherd{concurrently accessed} pages. Second, \emph{page-free} achieves almost the same performance as the \LocalOnly{} scheme. A small penalty is incurred as the first access to a page in \remoteMemory{} incurs %a 
cache line \shepherd{granularity latency access cost}. However, since the \shepherd{whole} page is migrated to \localMemory{} \emph{for free}, performance significantly improves thanks to spatial locality benefits of migrating pages in addition to the requested cache line. Thus, migrating pages to \localMemory{} is critical to \shepherd{achieving} high performance. Third, \emph{cache-line} outperforms \RemoteOnly{} in some latency-bound workloads with poor spatial locality, however its performance benefits depend on network characteristics. For example, in \emph{tr}, \emph{cache-line} outperforms \RemoteOnly{} by 1.42$\times$ when \shepherd{the} \netlat{} is 100 ns, while it incurs 1.82$\times$ performance slowdown over \RemoteOnly{} with 400 ns \netlat{}. Fourth, the \emph{cache-line+page} scheme, i.e., \emph{naively} moving data at \emph{both} \revised{granularities}, is still inefficient (only 1.11$\times$ better than \RemoteOnly), since critical cache lines are still queued behind large pages.

Overall, we draw two conclusions. (i) Page migrations incur high performance penalties and can significantly slow down the critical path cache line requests to other \shepherd{concurrently accessed} pages. However, if the overheads of migrating pages can be mitigated, \shepherd{moving data at page granularity} offers a critical opportunity to alleviate remote access costs. (ii) There is no \emph{one-size-fits-all} granularity in data movements to always perform best across all network configurations and applications. Depending on the spatial locality and the network load, some applications benefit from cache line-only accesses that avoid unnecessary congestion of pages in the network, while some applications significantly benefit from page movements that leverage spatial locality. \shepherd{To this end}, we design \myName{} to significantly reduce data movement costs across various \shepherd{application,} network and architecture characteristics. Figure ~\ref{fig:motivational-results} \shepherd{demonstrates} that \myName{} \shepherd{significantly} outperforms \shepherd{the} \RemoteOnly{} and \emph{cache-line+page} schemes by \shepherd{on average} 2.38$\times$ and 2.14$\times$, respectively.

%\vspace{-6pt}
\section{\myName: Our Approach}
%\vspace{-2pt}
\myName{} is an adaptive \revised{and scalable} data movement mechanism for \shepherd{fully} \disaggrSystems{} that supports low-overhead page migration, enables software transparency, and provides robustness to variations in \memoryComponent{} placements, network architectures and application behavior. 
%\revised{To enable independence and high scalability to multiple 
% a large pool of 
%\computeComponents{}/\memoryComponents{} that are flexibly added/removed in the \disaggrSystem, \myName{} enhances each \computeComponent{} and \memoryComponent{} with specialized engines, named \myName{} compute and memory engines, respectively (Fig.~\ref{fig:daemon-overview}).}  \revised{To achieve high system performance,} 
\myName{} comprises \fixed{two key ideas:}

\noindent\textbf{(1) Disaggregated Hardware Support for Data Movement Acceleration.} \fixed{We enhance each \computeComponent{} and \memoryComponent{} with specialized engines, i.e., \myName{} compute and memory engine (Figure~\ref{fig:daemon-overview}), respectively, to manage data \shepherd{movements} across the network of \disaggrSystems{}. 
\myName{} engines enable independence and high scalability to %multiple
\shepherd{a large number of} 
\computeComponents{}/\memoryComponents{} that are flexibly added/removed in \disaggrSystems. Moreover, distributed management of data migrations at multiple \myName{} engines increases the execution parallelism and decreases the processing costs and queuing delays to serve data requests.%, and better leverages the large aggregate network bandwidth enabled by multiple \memoryComponents{} of the system.
}

\noindent\textbf{(2) Synergy of Three Key Techniques.} \myName{} incorporates \fixed{three synergistic} \shepherd{key}
techniques shown in Figure~\ref{fig:daemon-overview}:

\begin{figure}[H]
    %\vspace{-8pt}
    \centering
    \includegraphics[width=1\linewidth]{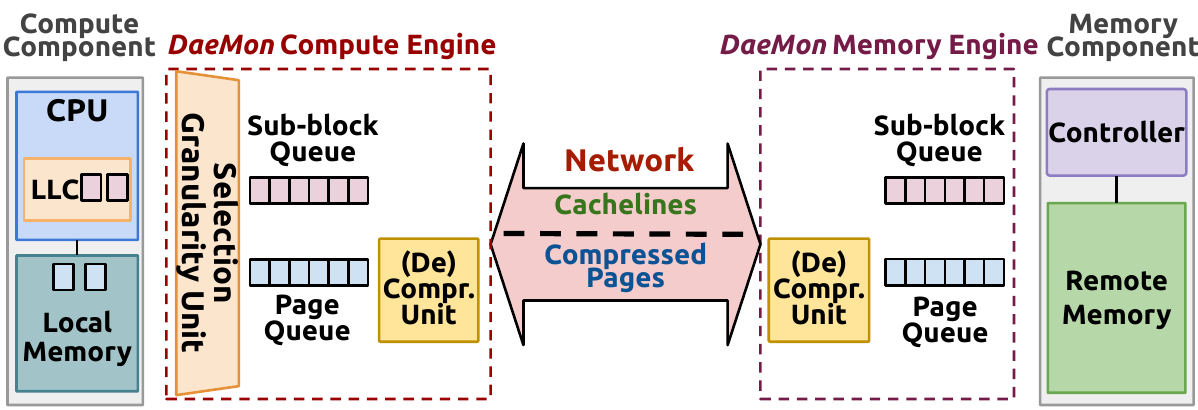}
    \vspace{-14pt}
    \caption{High-level overview of \myName.}
    \label{fig:daemon-overview}
    \vspace{-4pt}
\end{figure}

\noindent\textbf{\shepherd{(I)} \emph{Decoupled Multiple Granularity Data Movement}.} First, we integrate two separate hardware queues to manage and serve data requests from \remoteMemory{} at two granularities, i.e., cache line (via the sub-block queue) and page (via the page queue) granularity. Cache line requests are directly moved to Last Level Cache (LLC) \shepherd{of the \computeComponent{}} to avoid additional metadata overheads and eliminate memory latency.
%thus additional metadata is not required. 
Page \shepherd{requests} are moved to \localMemory{} \shepherd{of the \computeComponent{}}. Second, we \emph{prioritize} moving cache lines over moving pages via a bandwidth partitioning approach: a queue controller serves cache line and page requests with a \emph{predefined fixed ratio} to ensure that at any given time a certain fraction of the bandwidth resources is \emph{always allocated} to serve cache line requests quickly. \myName{} \shepherd{implements} both network and \remoteMemory{} bus bandwidth partitioning.

This technique provides two benefits. First, retaining page migrations \shepherd{in \myName{}} (i) enables software-transparency, (ii) allows maintaining metadata for DRAM at page granularity (thus incurring low metadata overheads), and (iii) exploits the performance benefits of data \shepherd{(spatial)} locality within pages. Second, cache line data movements that are on the critical path are quickly served, and have fewer slowdowns from expensive page \shepherd{movements} that may have been previously triggered, since \myName{} effectively prioritizes cache line movements.

\noindent\textbf{\shepherd{(II)} \emph{Selection Granularity Data Movement}.} To handle network, architecture and application variability in \disaggrSystems{}, we design a dynamic approach to decide whether a request should be served by a cache line, page, or \emph{both}, depending on application and network characteristics. 
At \myName{}'s engine of each \computeComponent{}, we include two separate hardware buffers to track \emph{pending} data migrations for both cache line and page granularity, and a selection granularity unit to control the granularity of upcoming data requests based on the utilization of the above buffers. The utilization of these buffers allows us to capture dynamic information regarding the current traffic in the system and the application behavior (i.e., locality). Our proposed selection granularity data movement enables robustness against fluctuations in network, architecture and application characteristics (we explain how this is implemented in §\ref{SelectingGranularitiesBl}).

\noindent\textbf{\shepherd{(III)} \emph{Link Compression on Page Movements}.} We leverage the decoupled page movement to use a \shepherd{high-latency} link compression scheme (with \shepherd{high compression ratio}),
%use a higher-latency link compression scheme (with higher compression ratios)
when \revised{moving} pages across the network. We integrate hardware compression units at both the \computeComponents{} and \memoryComponents{} to highly compress pages \revised{moved} over the network: \revised{the} page is compressed before it is being transferred over the network, and decompressed when it arrives at the destination (before it is written in memory \shepherd{modules}). \emph{Link compression} on page \shepherd{movements} reduces the network bandwidth consumption and alleviates network bottlenecks.

%\noindent\textbf{Synergy among \fixed{four} techniques.} 
Overall, \myName{} cooperatively integrates all three \shepherd{key} techniques, the synergy of which \shepherd{provides} a %more 
versatile solution: \\
%\myName{} cooperatively integrates all \fixed{four} techniques \revised{to provide a versatile solution:} \\
(1) Prioritizing requested cache lines helps \myName{} to tolerate high (de)compression latencies in page migrations over the network, while also leveraging benefits of page migrations (low metadata overheads, spatial locality). \\
(2) \shepherd{Moving compressed pages consumes} less network bandwidth, helping \myName{} to reserve part of the bandwidth to effectively prioritize critical path cache line accesses. \\
%(4) \revised{Compression on page moves}\cgiannou{is this correct? The Selection granularity mechanism helps daemon to adapt to data compressibility} 
(3) \fixed{Selection granularity movement} helps \myName{} to adapt to the \shepherd{application} data compressibility: if the pages are highly compressible, the number of \emph{pending} page migrations is relatively low, thus \myName{} favors \shepherd{moving data more at page granularity instead of cache line granularity} (and vice-versa).
%\myName{} favors sending more pages (and vice-versa).

%\vspace{-8pt}
\section{\myName: Detailed Design} 
%\vspace{-2pt}

\begin{figure}[t]
    %\vspace{-2pt}
    \centering
    \includegraphics[width=1\linewidth]{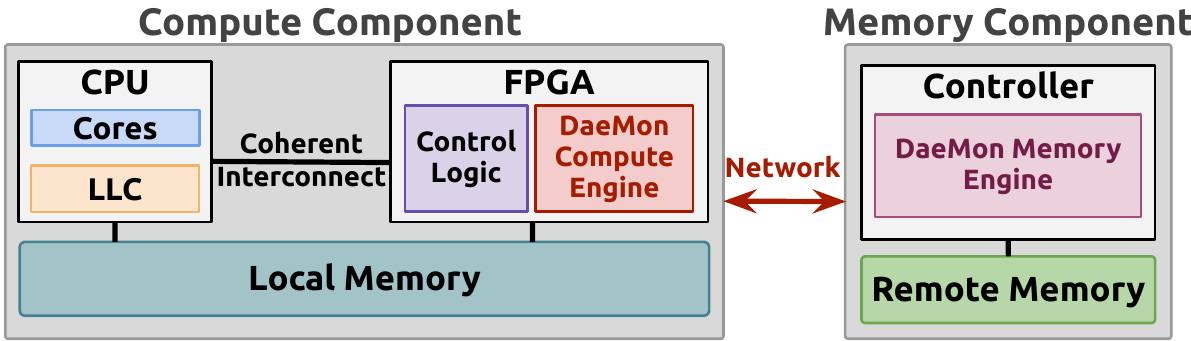}
    \vspace{-12pt}
    \caption{Proposed architecture \shepherd{for \computeComponent{} and \memoryComponent{}}.}
    \label{fig:reference-architecture}
    \vspace{-10pt}
\end{figure}

We design \myName{} to be a disaggregated solution:  a \myName{} compute engine is added at \emph{each} \computeComponent{} of the system to handle data requests to \remoteMemory{}, and a \myName{} memory engine is integrated at the controller of \emph{each} \memoryComponent{} of the system. Figure~\ref{fig:reference-architecture} shows our proposed architecture.

The \shepherd{baseline architecture of each \computeComponent{} includes a chiplet-based CPU+FPGA architecture (this CPU+FPGA integrated design has also been proposed to prior state-of-the-art work~\cite{Calciu2021Rethinking,Hwang2020Centaur}),} which is expected to have small cost~\cite{Calciu2021Rethinking} compared to the overall cost savings enabled by \disaggrSystems{}, while it is also 
socket compatible to current systems~\cite{Hwang2020Centaur,Cock2022Enzian}. The FPGA has three communication paths: i) a coherent path, i.e., CPU-FPGA coherent links, to access the CPU on-chip cache hierarchy, ii) an \revised{interface} (channel-based connection) 
to access the \localMemory{}, and iii) an external connection to the network controller to move data to/from \remoteMemory. We propose extending the FPGA by adding \shepherd{a new lightweight hardware component to handle data requests, i.e., the \myName{} compute engine.}

%The design is similar to state-of-the-art work~\cite{Calciu2021Rethinking,Hwang2020Centaur}: for each \computeComponent{} we include a chiplet-based CPU+FPGA architecture, which is expected to have small cost~\cite{Calciu2021Rethinking} compared to the overall cost savings enabled by \disaggrSystems{}, while it is also  socket compatible to current systems~\cite{Hwang2020Centaur,Cock2022Enzian}. The FPGA has three communication paths: i) a coherent path, i.e., CPU-FPGA coherent links, to access the CPU on-chip cache hierarchy, ii) an \revised{interface} %channel-based connection to access the \localMemory{}, and iii) an external connection to the network controller to move data to/from \remoteMemory. We propose extending the FPGA by adding \myName{} compute engine.%, as explained next.

Each \memoryComponent{} includes its own controller~\cite{Shan2018LegoOS,Lee2021MIND,guo2021clio}, that has two communication paths: a channel-based connection to DIMM modules \revised{of \remoteMemory}, and an external connection to the network, which is used to move data from/to \computeComponents{}. We propose extending \shepherd{the controller of each \memoryComponent{} by adding a new hardware component to handle data movements, i.e., the \myName{} memory engine.}
%We propose extending this controller by adding \myName{} memory engine.%, as described next. 

In our study, we assume that the \localMemory{} is an inclusive cache for the \remoteMemory{}, which contains all application data. The \localMemory{} implements an approximate LRU %cache
replacement policy, similar to prior state-of-the-art work~\cite{Shan2018LegoOS}. %Figure~\ref{fig:detailed_design} shows the detailed design of \myName{} compute and memory engine.

%\vspace{-1pt}
\subsection{Enabling \shepherd{Decoupled} Multiple Granularity Data Movement}\label{HandlingGranularitiesBl}
\begin{figure}[t]
    %\vspace{-1pt}
    \includegraphics[width=.74\textwidth]{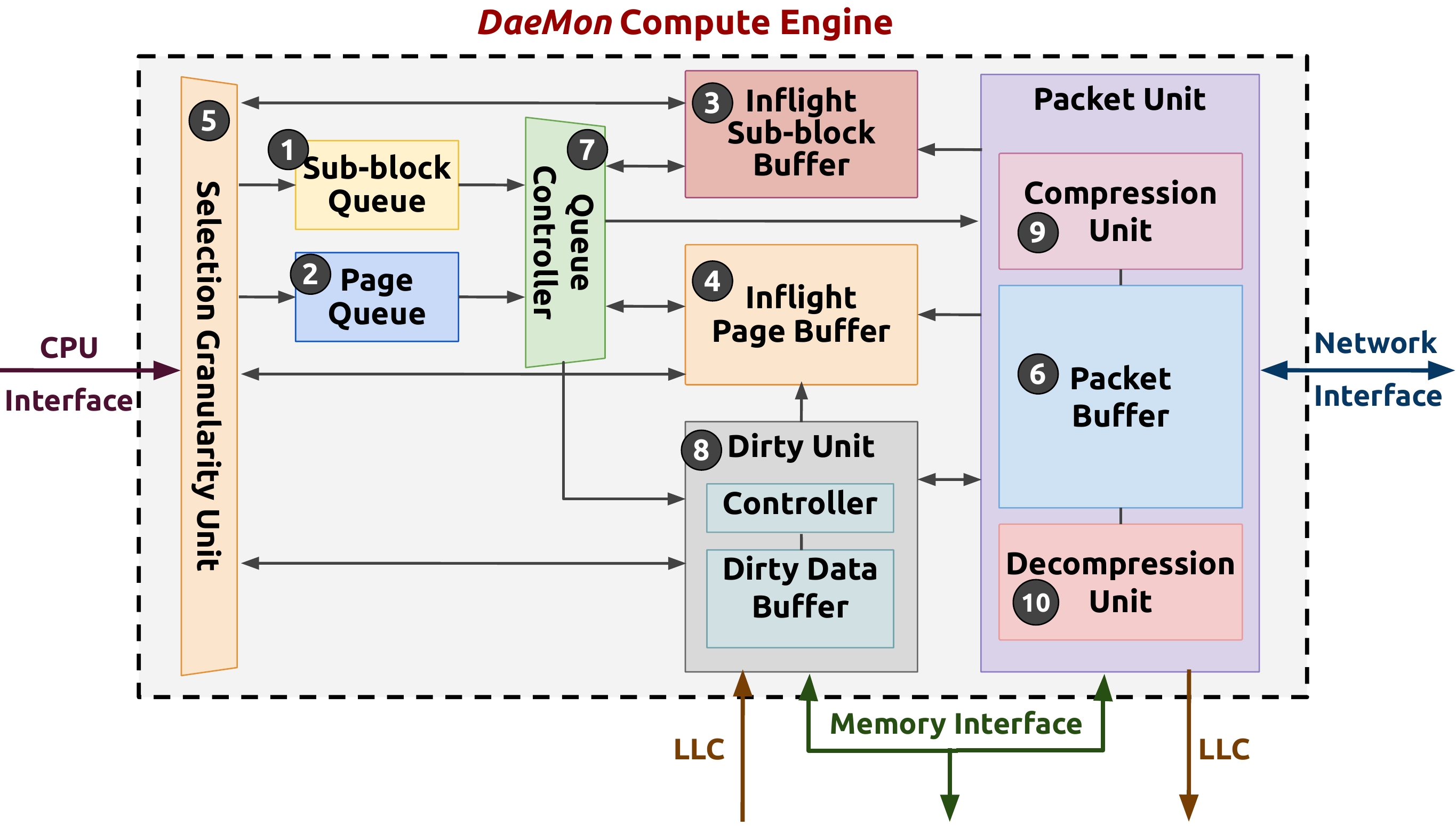}
    \includegraphics[width=.73\textwidth]{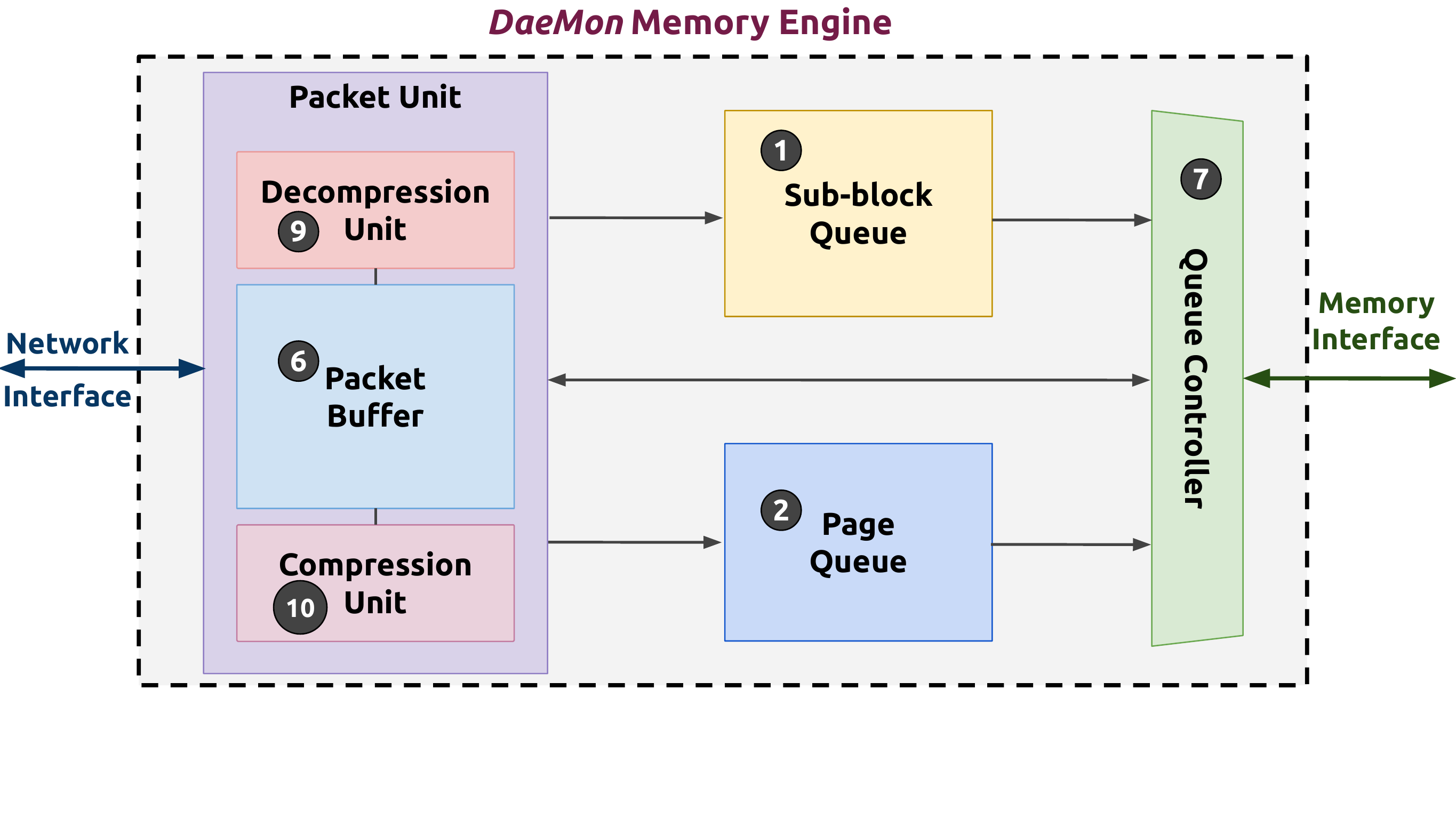}
    \vspace{-18pt}
    \caption{Detailed design of \myName{} engines for the compute (left) and memory (right) components. }
    \label{fig:detailed_design}
    \vspace{-6pt}
\end{figure}

%\vspace{-1pt}
Figure~\ref{fig:detailed_design} shows the detailed design of the \myName{} compute \shepherd{engine} and \myName{} memory engine. \myName{} engine includes two queues to handle requests at each granularity: cache line granularity via the sub-block queue \circled{1}, and page granularity via the page queue \circled{2}. It also includes a queue controller \circled{7} to serve requests from both queues, and a packet buffer \circled{6} to temporarily keep arrived packets, while they are being processed.

\noindent\textbf{Approximate Bandwidth Partitioning.} To prioritize cache line data movements while also ensuring that page movements are not aggressively stalled, we design an approximate bandwidth partitioning approach between the cache line and page movements, and configure the queue controller to serve cache line and page requests with a \emph{predefined fixed ratio}. Assuming that cache line and page requests transfer 64B and 4KB of data, respectively, and having a bandwidth partitioning ratio of 25\% (Figure~\ref{fig:single-pq} presents a sensitivity study on this ratio), 25\% of the bandwidth is reserved for cache lines as follows: for each page request issued through the network, which results in transferring 4KB data, the queue controller needs to serve $4096/64 * 0.25/(1-0.25) \approx 21$ cache line requests, each transferring 64B of data. To ensure this approximate partitioning is always maintained, we retain this alternate serving of page and cache line requests even if either queue is empty (i.e., requests may be not issued in all cycles). \myName{} implements an approximate bandwidth partitioning both in the network \shepherd{across components of the system} and when accessing data from \remoteMemory{} modules.

%\vspace{-1pt}
\subsection{Selecting the Data Movement Granularity}\label{SelectingGranularitiesBl}
%\vspace{-1pt}
\myName{} \shepherd{compute engine additionally} includes two separate hardware buffers to track data requests which are scheduled to be moved or in the process of being migrated (\shepherd{henceforth} referred to as \textit{inflight}): (i) the inflight sub-block buffer for the cache line granularity requests \circled{3}, and (ii) the inflight page buffer for the page granularity requests \circled{4}. Both buffers are used to track \emph{pending} data migrations and avoid requesting the same data multiple times. \myName{} \shepherd{compute engine} includes a selection granularity unit \circled{5} which throttles data requests to avoid requesting the same data multiple times, and decides at which granularity the request should be served (cache line, page, or both \shepherd{granularities}).

\noindent\textbf{Scheduling Page Granularity Data Movements.}
When the \myName{} \shepherd{compute engine} receives a data request, the selection granularity unit checks (i) the utilization of the inflight page buffer, and (ii) if the corresponding page has already been scheduled to be moved. If the page has already been requested or the inflight page buffer is full, the selection granularity unit \emph{does not request the page}. Thus at any given time, the number of pages scheduled to be moved is automatically limited by the selection granularity unit, also limiting storage/area overheads to track the pending page migrations.

If the inflight page buffer is not full, \shepherd{the selection granularity units schedules} the page migration by adding a new entry in the page queue and the inflight page buffer, marking the page as \textit{scheduled}. When the queue controller issues the movement, the corresponding entry is released in the page queue, and the page entry in the inflight page buffer is marked as \textit{moved}. When the requested page arrives, the corresponding entry is released (\textit{invalid} state) in the inflight page buffer. The page is written to \localMemory{} and all pending requests are serviced via \localMemory{}. Any entries in the inflight sub-block buffer with requests to cache lines in the same page are removed and thus, any data packets that arrive in the future with cache lines from the same page are simply ignored. \revised{In} \myName{}, we retain existing data management and address translation mechanisms at page granularity. Local page table updates at the \computeComponent{} are only performed \revised{in page migrations.}% when a page is migrated.

\noindent\textbf{Scheduling Cache Line Granularity Data Movements.}
To decide whether a cache line granularity movement should be made, the selection granularity unit checks (i) the utilization of the inflight sub-block buffer and (ii) if the corresponding page was already scheduled to be moved (by a previous request). There are two cases. First, if the corresponding page is \emph{not} scheduled to be moved according to the inflight page buffer, the selection granularity unit always schedules a cache line granularity data movement. Second, if the corresponding page is already scheduled to be moved, the selection granularity unit sends the cache line only if: (i) the sub-block buffer has lower utilization than the page buffer and (ii) the page is not already in the process of migration (i.e., the page is in the page queue). Otherwise, it drops the request as the page has already been requested. This avoids unnecessarily sending cache lines when the corresponding page is likely to arrive faster and when the sub-block queue is likely to be slow due to oversaturation.

If a cache line is scheduled, a new entry is added both in the sub-block queue and the inflight sub-block buffer. When the queue controller issues the movement, the corresponding entry is released in the sub-block queue. When the requested cache line arrives at the \computeComponent{}, the corresponding entry %for the requested cache line 
is released in the sub-block buffer, and the data is \emph{directly} written to LLC through the FPGA-based coherent interconnect.

The above mechanism enables an adaptive approach for the data movement granularity based on \revised{the dynamic} network/architecture and application characteristics:

\noindent(1) If there is \emph{high locality} within pages, there are fewer pages requested, and the sub-block buffer fills up faster than the page buffer. Thus, \myName{} favors issuing pages and throttles cache line requests. If there is \emph{low  locality} within pages, the page buffer fills up faster than the sub-block buffer, since  cache line requests are served at a higher rate than page requests (e.g., 21:1 cache lines versus pages requests 
for 25\% bandwidth ratio). Thus, \myName{} favors issuing cache line \shepherd{movements} and throttles page \shepherd{migrations}.

\noindent(2) \rebuttal{If both the page and  sub-block buffers are fully utilized, \myName{} detects \emph{bandwidth constrained} scenarios. In bandwidth constrained scenarios, \myName{} favors issuing more cache line \shepherd{movements} to alleviate bandwidth bottlenecks. When the bottleneck is mitigated, (inflight buffers are not fully utilized), \myName{} schedules more page \shepherd{movements} to obtain locality benefits.}

%\noindent(2) Under \emph{bandwidth constrained} scenarios \myName{} favors issuing more cache line data movements, as the inflight page buffer quickly saturates and thus pages are throttled. Instead, under less bandwidth constrained scenarios, \myName{} schedules more pages to obtain locality benefits. 

\noindent(3) Additionally, when using link compression to transfer pages, \myName{} is able to adapt to the compressibility of \shepherd{the} application data: if the pages are highly compressible, the inflight page buffer empties at a faster rate and thus \myName{} favors sending more page \shepherd{migrations} (and vice versa).

%\vspace{-1pt}
\subsection{Handling Dirty Data} 
%\vspace{-1pt}
\rebuttal{Dirty data (cache lines/pages) is always directly written to \remoteMemory.}
Data (cache line or page granularity) can be in one of the three states: (i) \textit{local:} when data is cached in on-chip caches \shepherd{(for cache lines)} or \localMemory{} \shepherd{(for pages)}, (ii) \textit{remote:} when data is only in \remoteMemory{}, and (iii) \textit{inflight:} when data is being migrated. With \myName{}, data can be present simultaneously in two states: for example, local as a cache line (in the cache hierarchy \shepherd{of the compute component}) and inflight as a page or vice versa. This poses coherence issues if the processor writes to data in the above state.

There are two scenarios: (i) if a page arrives to \computeComponent{} \emph{before} a prioritized cache line, any modifications to the page may be overwritten by the stale cache line that arrives later, and (ii) if a dirty cache line is evicted from the LLC while the corresponding page is in transit, %(or in \remoteMemory), 
the modifications would be lost when the page arrives to \computeComponent{}. As explained, in the (i) scenario, when a page arrives, the corresponding entries in the inflight sub-block buffer with requests to cache lines in the same page are removed and thus, any data packets that arrive in the future with cache lines from the same page are simply ignored. In the (ii) scenario, for every dirty cache line that gets evicted by the LLC and also misses in the \localMemory, its corresponding page can be either inflight or in \remoteMemory. To ensure correctness, \myName{} \shepherd{compute engine} first checks if there is an inflight page request in the inflight page buffer. If there is \emph{no} inflight page request (\shepherd{according to the inflight page buffer)}, the evicted dirty cache line is directly migrated to \remoteMemory{}. In the other case, we need to retain the dirty cache line until the page arrives. We include a dirty unit \circled{8} in \shepherd{the \myName{} compute engine} with a \emph{dirty data buffer} that temporarily stores these dirty cache lines. When the corresponding inflight page arrives, \shepherd{the \myName{} compute engine} flushes the dirty cache line(s) from the dirty buffer to \localMemory.

Prior works~\cite{Calciu2021Rethinking,Adya2019Fast} observe that typically a few cache lines (1-8 cache lines) or all cache lines of a page are accessed. Thus, when the evicted dirty cache lines of the same page increase beyond a predefined threshold (e.g., 8 cache lines), \shepherd{the \myName{} compute engine} flushes all dirty cache lines to \remoteMemory, and marks the corresponding entry for that page in the inflight page buffer as \emph{throttled}. When the inflight page arrives, \shepherd{the \myName{} compute engine} ignores it, since its entry is in the \emph{throttled} state, and sends a new request for that page to receive the up-to-date data. This enables lower area/storage overheads for the dirty data buffer.

%\vspace{-1pt}
\subsection{Link Compression in Page Migrations}\label{CompressionBl} 
%\vspace{-1pt}

Approaches for data compression are typically of two types: (i) latency-optimized compression 
schemes~\cite{Pekhimenko2012BDI,Alameldeen2004FPC,Chen2010CPack,Yang2000Frequent,Yang2004Frequent}, which optimize\shepherd{/minimize} the (de)compression latencies, and (ii)  
ratio-optimized compression schemes~\cite{Abali2001MXT,Tremaine2001MXT,Kim2017Transparent,Ziv1977LZ}, which provide higher compression ratios while incurring relatively high (de)compression latencies. We select a ratio-optimized compression scheme in \myName{} based on two observations (§\ref{Evaluationbl}): (i) in \disaggrSystems{},  queueing delays and network latencies can be significant, thus %the 
compression benefits outweigh the high (de)compression latencies, and (ii)  \myName{} prioritizes cache lines that are on the critical path, thus we can tolerate relatively high (de)compression latencies for page migrations.

\myName{} engines include (de)compression units \circled{9} \circled{10} that compress pages transferred through the network. We implement a hardware design similar to IBM MXT~\cite{Abali2001MXT,Tremaine2001MXT}, using the LZ77 compression algorithm~\cite{Ziv1977LZ}, and operating at 1KB granularity at a time. (De)Compression units include 4 engines, each of which operates on 256B of data and uses a 256B shared dictionary, incurring in total a 64-cycle latency according to~\cite{Abali2001MXT,Tremaine2001MXT}.

%\vspace{-1pt}
\subsection{\myName's Hardware Structures}
%\vspace{-1pt}

\begin{figure}[H]
    \vspace{-5pt}
    \centering
    \includegraphics[width=\linewidth]{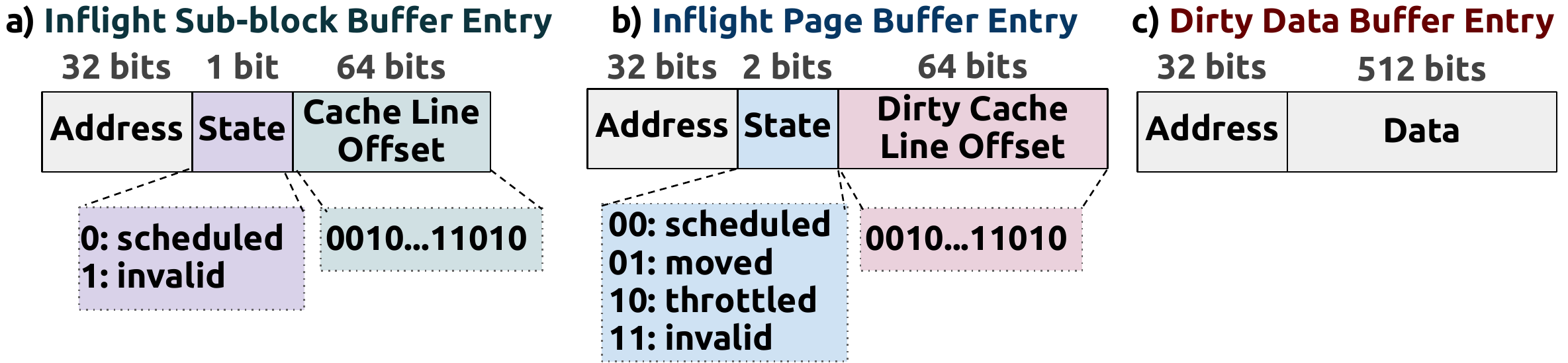}
    \vspace{-12pt}
    \caption{An inflight sub-block buffer entry, an inflight page buffer entry, and a dirty data buffer entry.}
    \label{fig:buffer_entries}
    \vspace{0pt}
\end{figure}

We estimate the overheads of \myName{}'s hardware structures for each \computeComponent{} assuming a 64-core CPU, using CACTI~\cite{Muralimanohar2007Optimizing}. \shepherd{The sizes of the \myName{} sub-block and page queues and the sub-block and page buffers have been selected based on the maximum possible number of \emph{pending} data migrations at a time, which is determined by the number of the available LLC MSHRs (Miss Status Holding
Registers) in a typical CPU system, and is independent of the workloads’ patterns
and the mix of workloads that are running at each time.} For the hardware structures at each \memoryComponent{}, we scale \shepherd{the sizes of the \myName{} sub-block and page queues}, assuming that each \memoryComponent{} can concurrently serve up to 4 \computeComponents{}. Table~\ref{table:hardware_overhead} \shepherd{ presents the hardware overheads of \myName{} compute engine (C) and \myName{} memory engine (M).} Figure~\ref{fig:buffer_entries} shows an inflight sub-block buffer entry, an inflight page buffer entry, and a dirty data buffer entry.

\begin{table}[H]
    %\vspace{-14pt}
    \centering
    \resizebox{0.86\columnwidth}{!}{%
    %\begin{threeparttable}
    \begin{tabular}{l r r c c c}
    \toprule
    \textbf{Hardware} & \textbf{Entries} & \textbf{Size} & \textbf{Access} & \textbf{Area } & \textbf{Energy}
    \\
    \textbf{Structure} & \textbf{} & \textbf{(KB)} & \textbf{Cost (ns)} & \textbf{Cost (mm\textsuperscript{2})} & \textbf{Cost (nJ)}
    \\    
    \hline
    Sub-block Queue (C) & 128 & 0.5 & 0.34 & 0.084 & 0.038 \\
    \hline
    Sub-block Queue (M) & 512 & 2 & 0.38 & 0.093 & 0.039 \\
    \hline    
    Page Queue (C) & 256 & 1 & 0.35 & 0.087 & 0.038 \\
    \hline
    Page Queue (M) & 1024 & 4 & 0.40 & 0.105 & 0.041 \\
    \hline  
    Inflight Sub-block Buffer (C) & 128 & 1.625 & 0.56 & 0.041 & 0.046 \\
    \hline    
    Inflight Page Buffer (C) & 256 & 3.25 & 0.77 & 0.089 & 0.096 \\
    \hline       
    Dirty Data Buffer (C) & 256 & 17 & 0.62 & 0.168 & 0.046 \\
    \hline  
    Packet Buffer (C) & - & 8 & 0.538 & 0.137 & 0.044 \\
    \hline   
    Packet Buffer (M) & - & 32 & 1.032 & 0.263 & 0.047 \\
    \hline    
    2 $\times$ Dictionary Table (C,M) & 1024 & 1 & 0.28 & 0.015 & 0.020 \\
    \hline       
    \bottomrule
    \end{tabular}
 }
 \vspace{6pt}
 \caption{\myName's hardware overheads for C: compute engine and M: memory engine.}
\label{table:hardware_overhead}
\vspace{-12pt}
\end{table}

\noindent\textbf{Sub-block Queue (SRAM), 128 entries}: The sub-block queue \revised{size} is limited by the available LLC MSHRs of the \computeComponent{}. \\
\noindent\textbf{Page Queue (SRAM) - 256 entries}: The page queue has 256 entries, since \myName{} serves requests from the page queue at a smaller rate than the sub-block queue.\\ 
\noindent\textbf{Inflight Sub-block Buffer (CAM) - 128 entries}: Similar to the sub-block queue, \fixed{this buffer has} 128 entries. We design this hardware structure to be indexed using the corresponding page address to achieve smaller area costs, since at a given time there may be multiple inflight cache line requests to the same page. Each entry (Figure~\ref{fig:buffer_entries}a) includes the page address, the state (\emph{scheduled} or \emph{invalid}), and a 64-bit queue that is used to indicate the offsets within the page of the inflight cache requests by (re)setting the corresponding bits. \\
\noindent\textbf{Inflight Page Buffer (CAM) - 256 entries}: An inflight page buffer entry (Figure~\ref{fig:buffer_entries}b) includes the page address, the state that can be \emph{scheduled}, \emph{moved}, \emph{throttled} (when the page needs to be re-requested) or \emph{invalid}, and a 64-bit queue to indicate the offsets of the dirty cache lines of the inflight page that are temporarily kept in the dirty data buffer.\\
\noindent\textbf{Dirty Data Buffer (SRAM) - 256 entries}: A dirty data buffer entry (Figure~\ref{fig:buffer_entries}c) includes the evicted cache line and its address.\\
\noindent\textbf{Packet Buffer (SRAM) - 8KB}: We use an 8KB buffer to temporarily store arrived data packets until they are processed. \\
\noindent\textbf{Dictionary Tables for (De)Compression (CAM) - 2KB}: \myName{} proposes 4 engines at each (de)compression unit, each of them has 256B CAM~\cite{Abali2001MXT,Tremaine2001MXT}. In total, we estimate each dictionary table as 1KB CAM.

Overall, \shepherd{\myName{}'s hardware overheads are due to the cache memories corresponding to the sub-block and page queues, the sub-block and page buffers, and the dictionary tables used for data compression. The total sizes of the \myName{} cache memories are $\sim$34KB and 40KB for the \myName{} compute and memory engine, respectively. Therefore, \myName{}'s hardware overheads are similar to that of the small L1 cache memory of a modern state-of-the-art processor (e.g., Intel Xeon). We conclude that our
proposed hardware structures incur very modest hardware and financial costs to be integrated into the \computeComponents{} and \memoryComponents{} of \disaggrSystems{}. }

%\vspace{-1pt}
\subsection{Handling Failures}
%\vspace{-1pt}

\myName{} handles \computeComponent{}, \memoryComponent{} and network failures using fault-tolerance approaches of prior works~\cite{Lee2021MIND,Shan2018LegoOS,Calciu2021Rethinking}. If the \computeComponent{} fails (CPU or \myName{} compute engine), the application needs to be restarted potentially on a different \computeComponent{} of the system. Network failures are handled using timeouts: %the 
\myName{} engines can trigger timeouts when pending page or cache line requests have not arrived after a long time, or when ACK messages have not been received for migrations of dirty data. \shepherd{The exploration of the timeout period value is left for future work.} Finally, \memoryComponent{} failures are handled via data replication, \shepherd{similarly to prior work}~\cite{Shan2018LegoOS}: \myName{} can send the evicted dirty data to more than one \memoryComponent{}, and wait to receive ACK messages from all of them.

%\vspace{-1pt}
\subsection{\myName{} Extensions}\label{ExtensionsBl} 
%\vspace{-1pt}
%\noindent\textbf{Prefetching.} \myName{} can be flexibly extended to support hardware/software-based prefetchers. Existing CPU prefetchers might generate additional data requests, which are handled by \myName{} normally and are served by migrating the prefetched data at a cache line, page granularity or both, \revised{via} on our \revised{proposed} selection granularity scheme.\\ 
\noindent\textbf{Prefetching.} \myName{} \rebuttal{can flexibly support hardware/software-based prefetchers. Existing CPU prefetchers might generate data requests, which  \myName{} can normally serve by migrating the prefetched data at a cache line granularity, page granularity or both granularities, via on our proposed selection granularity scheme. Page prefetchers~\cite{Maruf2020Effectively} might generate page-granularity data requests, which \myName{} can serve by migrating the prefetched data at page granularity or throttling the page request based on our selection granularity scheme.}\\ 
\noindent\textbf{Large Pages.} \myName{} can be easily extended to support large granularity pages (e.g., 2MB). To effectively prioritize cache line requests over page requests, \myName's predefined ratio for the approximate bandwidth partitioning needs to be properly configured based on the size of the large page. \rebuttal{To enable multiple page sizes (e.g., both 4KB and 2MB), we could enhance \myName{} to split large pages (e.g., 2MB) to consecutive page requests of smaller sizes (e.g., 4KB) issued in the page queue. }

%\vspace{-8pt}
\section{Methodology}\label{Methodologybl}
%\vspace{-3pt}

\noindent\textbf{Simulation Methodology.} 
We use Sniper~\cite{carlson2011etloafsaapms,carlson2014aeohmcm}, \shepherd{a state-of-the-art accurate simulator, and we heavily modified it to} model a \disaggrSystem{} with one \computeComponent{} and multiple \memoryComponents{} interconnected across the network. We present detailed evaluation results using one \memoryComponent{} and provide 
a characterization study of multiple \memoryComponents{} with various network configurations in Figure~\ref{fig:multimem}. For the network across components, we  \revised{use both (i) a fixed latency of} 100 ns/400 ns~\cite{Gao2016Network,Lee2021MIND} to model propagation and switching delays inside network (referred to as \emph{\netlat}), and \revised{(ii) a variable latency of modeling  
the current bandwidth utilization at each simulation interval (100K ns) when configuring} the network bandwidth to be 2-8$\times$ less than DRAM bandwidth~\cite{Shan2018LegoOS,Gao2016Network} (referred to as \emph{\bwf}). \shepherd{For the compute component, we configure a state-of-the-art CPU server with on-chip cache memories of typical sizes and x86 OoO cores of 3.6GHz frequency.  The \localMemory{} size is configured to fit $\sim$20\% of each application's working set, and we evaluate LRU replacement policy~\cite{Shan2018LegoOS} in \localMemory{}, unless otherwise stated. The aforementioned configuration is
consistent with prior state-of-the-art works in disaggregated systems~\cite{Shan2018LegoOS,Lee2021MIND,Gao2016Network}. For both the \localMemory{} and \remoteMemory{}, we evaluate a DDR4 memory model with 17GB/s bus bandwidth, and we simulate hardware-based address translation for memory pages, having overhead as one DRAM access cost per lookup, as explained in prior state-of-the-art work~\cite{guo2021clio}. We evaluate access overheads in \myName{} queues/buffers using CACTI~\cite{Muralimanohar2007Optimizing} (See Table~\ref{table:hardware_overhead}). Table~\ref{table:sniper_parameters} lists the parameters of our simulated system.}

%We simulate hardware-based address translation having overhead as one DRAM access cost, as explained in prior state-of-the-art work~\cite{guo2021clio}. Table~\ref{table:sniper_parameters} lists the parameters of our simulated system. \revised{We evaluate LRU replacement policy~\cite{Shan2018LegoOS} in \localMemory{}, unless otherwise stated.}

\begin{table}[H]
    %\vspace{-2pt}
    \centering
    \resizebox{0.94\columnwidth}{!}{%
    \begin{tabular}{l l}
    \toprule
    \textbf{CPU} & 3.6 GHz, 4-way OoO x86 cores, \revised{224-entry ROB;}
    \\
    \hline
    \textbf{L1 Instr. Cache} & 32 KB, 4-way associativity, LRU; \\
    \hline
    \textbf{L1 Data Cache} &32 KB, 8-way associativity, 4-cycle access latency, LRU; \\  
    \hline
    \textbf{L2 Cache} & 256 KB, 8-way associativity, 8-cycle access latency, LRU; \\
    \hline
    \textbf{LLC} &  4MB, 16-way associativity, 30-cycle access latency, LRU;  \\
    \hline    
    \textbf{Local Memory} & 2400MHz, 15ns process. latency, 17GB/s bus bandwidth ~\cite{ddr4_2400}; \\
    \hline  
    \textbf{Network} & 2-8$\times$ less than bus bandwidth, 100-400 ns switching latency~\cite{Gao2016Network,Shan2018LegoOS}; \\   %~\cite{Gulur2015acomprehensive} for queueing latency; \\
    \hline  
    \textbf{Remote Memory} & 2400MHz, 15ns process. latency, 17GB/s bus bandwidth ~\cite{ddr4_2400}; \\    
    \bottomrule
    \end{tabular}
 }
 \vspace{4pt}
 \caption{Configuration of simulated system.}
\label{table:sniper_parameters}
\vspace{-12pt}
\end{table}

\noindent\textbf{Workloads.} We evaluate \fixed{various} workloads with different memory access patterns from \shepherd{various application domains including} graph processing, machine learning, bioinformatics, linear algebra, data analytics, and HPC domains, \shepherd{shown in} Table~\ref{table:workloads}. \fixed{The dynamic working sets at any given point at runtime range from 43.2MB to 1.32GB. In a fully \disaggrSystem{}, the application working set (irrespective of the size) is primarily housed in \remoteMemory{} to provide the benefits of improved elasticity, heterogeneity, and failure isolation. Therefore, we configure the \localMemory{} size to fit $\sim$20\% of each application's working set (similar to \shepherd{prior state-of-the-art work}~\cite{Shan2018LegoOS,Lee2021MIND,Gao2016Network}). All data is initially located in \remoteMemory{}.}
%Even for small working sets data is hosted by \remoteMemory{}~\cite{guo2021clio,Shan2018LegoOS,Lee2021MIND,Gao2016Network}, and \disaggrSystems{} can still provide high cost benefits by improving elasticity, heterogeneity, and failure isolation.} %\fixed{device failure handling, and resource scaling.}
%\fixed{\disaggrSystems{}  high cost benefits by \emph{fully} disaggregating memory from compute units, thus even small working sets are allocated and hosted by \remoteMemory{}~\cite{guo2021clio,Shan2018LegoOS,Lee2021MIND,Gao2016Network}.} 
%We configure the \localMemory{} size to fit $\sim$20\% of each application's working set (similar to~\cite{Shan2018LegoOS,Lee2021MIND,Gao2016Network}), and all data is initially located at \remoteMemory{}. 
We simulate most workloads to full execution and for slower long running workloads, we simulate 1B instructions.  

%\noindent\textbf{Workloads.} We evaluate various workloads with different memory access patterns from graph processing, machine learning, bioinformatics, data analytics, and HPC domains (Table~\ref{table:workloads}). We simulate most workloads to full execution and for slower long running workloads, we simulate %up to 
%1B instructions. The \localMemory{} size is configured to fit $\sim$20\% of each application's working set, while \remoteMemory{} hosts all remaining data (similar to~\cite{Shan2018LegoOS,Lee2021MIND,Gao2016Network}). All data is initially located at \remoteMemory{}.

%\begin{comment}
\begin{table}[!htb]
\begin{minipage}{0.94\linewidth}
  %\vspace{-2pt}
   %\hspace{-4pt}
   \centering
   \resizebox{\textwidth}{!}{
    \begin{tabular}{l l l c} 
    \toprule
    \textbf{Workload} & \textbf{Domain} &  \textbf{Input Data}  \\ %[0.5ex] 
    \midrule
    \midrule
    K-Core Decomposition (\textbf{kc})~\cite{Ligra2013Ligra} & Graph Processing & 1M vertices x 10M edges  \\ 
    \hline    
    Triangle Counting (\textbf{tr})~\cite{Ligra2013Ligra} & Graph Processing &  1M vertices x 10M edges \\
    \hline    
    Page Rank (\textbf{pr})~\cite{Ligra2013Ligra} & Graph Processing & 1M vertices x 10M edges  \\
    \hline     
    Needle Wunsch (\textbf{nw})~\cite{Che2009Rodinia} & Bioinformatics & 4096 base pairs per sequence \\
    \hline     
    Breath First Search (\textbf{bf})~\cite{Ligra2013Ligra} & Graph Processing & 1M vertices x 10M edges \\
    \hline
    Betweenness Centrality (\textbf{bc})~\cite{Ligra2013Ligra} & Graph Processing & 1M vertices x 10M edges  \\
    \hline
    Timeseries (\textbf{ts})~\cite{MPROFILEI} & Data Analytics & 262144 elements in sequence \\     
    \hline 
    Sparse Matrix Vector Multipl. (\textbf{sp})~\cite{Kjolstad2017Taco} & Linear Algebra & pkustk14 matrix \\
    \hline
    Sparse Lengths Sum (\textbf{sl})~\cite{NaumovDLRM19} & Machine Learning & Kaggle Criteo 10GB Dataset \\
    \hline
    High Perf. Conjugate Gradient (\textbf{hp})~\cite{hpcg} & HPC & 104 x 104 x 104 \\
    \hline
    Particle Filter (\textbf{pf})~\cite{Che2009Rodinia} & HPC & 4096 x 4096, 30000 particles\\    
    \hline
    Darknet19 (\textbf{dr})~\cite{darknet13} & Machine Learning & dog.jpg (768 x 576 pixels) \\ 
    \hline
    Resnet50 (\textbf{rs})~\cite{darknet13} & Machine Learning & dog.jpg (768 x 576 pixels)  \\        
    \hline   
    \bottomrule
    \end{tabular}}
  \end{minipage}
  \vspace{6pt}
  \caption{\label{table:workloads} Summary of  workloads.}
  %\vspace{-20pt}
\end{table}
%\end{comment}
%\vspace{-4pt}
\section{Evaluation}\label{Evaluationbl}
%\vspace{-1pt}

We evaluate six schemes: (i) \textbf{\RemoteOnly}: the typically-used approach~\cite{Shan2018LegoOS,Lee2021MIND,Calciu2021Rethinking} of \revised{moving} data to/from \remoteMemory{} at page granularity; (ii) \textbf{\LC}: \myName{}'s link compression for page movement without \shepherd{enabling} cache line granularity data movement, \shepherd{i.e., moving data at page granularity with LZ-based link compression enabled}; (iii) \textbf{\BP}: enabling only \myName{}'s decoupled multiple granularity data movement with 25\% bandwidth partitioning ratio for cache line \shepherd{movements}, i.e., moving data \emph{always} at both granularities; (iv) \textbf{\PQ}: enabling \myName{}'s \emph{both} decoupled multiple granularity and selection granularity data movement with 25\% bandwidth partitioning ratio for cache line \shepherd{movements} (without \shepherd{enabling data} compression in page migrations); (v) \textbf{\myName{}}: \myName{}'s complete design enabling all its three techniques (25\% bandwidth partitioning ratio); and (vi) \textbf{\LocalOnly}: the monolithic approach where all the data fits in \localMemory{} \shepherd{of the \computeComponent{}.}

%\vspace{-2pt}
\subsection{Performance}
\begin{figure}[t]
    \vspace{3pt}
    \centering
    \includegraphics[width=0.84\linewidth]{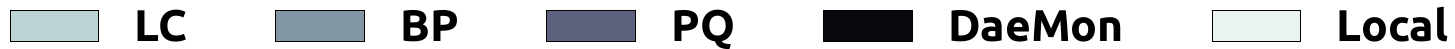}\hspace{200pt}
    \includegraphics[width=\linewidth]{results/singlethread_overview_netlat_100_bwsf_2_speedup.pdf}
    \includegraphics[width=\linewidth]{results/singlethread_overview_netlat_100_bwsf_4_speedup.pdf}
    \includegraphics[width=\linewidth]{results/singlethread_overview_netlat_100_bwsf_8_speedup.pdf}
    \includegraphics[width=\linewidth]{results/singlethread_overview_netlat_400_bwsf_2_speedup.pdf}
    \includegraphics[width=\linewidth]{results/singlethread_overview_netlat_400_bwsf_4_speedup.pdf}
    \includegraphics[width=\linewidth]{results/singlethread_overview_netlat_400_bwsf_8_speedup.pdf}    
    \vspace{-16pt}
    \caption{Speedup in \revised{all workloads} normalized to \RemoteOnly{} \revised{using various network configurations}.}
    \label{fig:single-performance}
    \vspace{-8pt}
\end{figure}

%\vspace{-2pt}
%\noindent\textbf{Performance.}  
Figure~\ref{fig:single-performance} compares all schemes with different network configurations. Our evaluated workloads exhibit three patterns and we make the following observations.

\begin{comment}
\begin{figure}[H]
    \vspace{-6pt}
    \centering
    \includegraphics[width=0.64\linewidth]{results/legend_speedup.png}\hspace{200pt}
    \includegraphics[width=0.495\linewidth]{results/singlethread_overview_netlat_100_bwsf_2_speedup.pdf}
    \includegraphics[width=0.495\linewidth]{results/singlethread_overview_netlat_400_bwsf_2_speedup.pdf}
    
    \includegraphics[width=0.495\linewidth]{results/singlethread_overview_netlat_100_bwsf_4_speedup.pdf}
    \includegraphics[width=0.495\linewidth]{results/singlethread_overview_netlat_400_bwsf_4_speedup.pdf}
    
    \includegraphics[width=0.495\linewidth]{results/singlethread_overview_netlat_100_bwsf_8_speedup.pdf}
    \includegraphics[width=0.495\linewidth]{results/singlethread_overview_netlat_400_bwsf_8_speedup.pdf}    
    \vspace{-20pt}
    \caption{Speedup in \revised{all workloads} normalized to \RemoteOnly{} \revised{using various network configurations}.}
    \label{fig:single-performance}
    \vspace{-12pt}
\end{figure}
\end{comment}

First, \emph{kc, tr, pr, and nw} exhibit relatively poor spatial locality within pages. In such workloads, \BP{} effectively prioritizes critical cache line requests. However, \PQ{} provides significant benefits thanks to dynamically selecting the \shepherd{data movement} granularity: the page buffer saturates faster than the sub-block buffer given the poor locality and the higher servicing rate of the cache line requests in the queue controller, thus the selection granularity unit enables the movement of more cache lines and fewer pages. This \shepherd{results} in reduced access latencies as critical path cache line \shepherd{requests} are no longer stalled behind many page migrations.

Second, \emph{bf, bc, and ts} exhibit medium spatial locality within pages. In such workloads, both \LC{} and \PQ{} decrease data access costs using different approaches: \LC{} enables exploiting more spatial locality by moving more pages, while \PQ{} accelerates accesses to the critical path cache line \shepherd{requests}, both of which benefit these workloads. 

Third, the remaining workloads exhibit high spatial locality within pages, thus page migration is critical to leverage data locality. In these workloads,
\BP{} incurs high performance slowdowns, since it is \emph{oblivious} to application behavior. Instead, \PQ{} effectively enables more page \shepherd{movements} and throttles cache line \shepherd{movements} by tracking pending data requests, thus \shepherd{achieving similar system performance} to \RemoteOnly{}.
\LC{} performs better for \emph{sp, sl, hp, and pf}, \shepherd{since these workloads} %they 
have higher data compressibility than \emph{dr} and \emph{rs}.

Fourth, when \shepherd{network} bandwidth is more constrained, \LC{} provides \emph{even higher} performance over \RemoteOnly{}, while \PQ{} is %largely 
unaffected by bandwidth as the bandwidth partitioning approach prioritizes cache line \shepherd{movements} even with low available bandwidth.

Fifth, \PQ{} is slightly affected by the \netlat{} (\shepherd{Please also see Figure~\ref{fig:netlat-sens} in Appendix §~\ref{Appendixbl}}): %e.g., 
\PQ{} outperforms \RemoteOnly{} by 1.60$\times$ and 1.51$\times$ for 100 ns and 400 ns \netlat{}, respectively. The slightly lower benefits are due to \PQ{}'s inability to hide %fixed 
network switch latencies in critical cache line movements. Instead, \LC{} is unaffected by \netlat{}, as page movement incurs much higher overheads \shepherd{(due to very high network processing and queueing delays)} over the smaller \netlat{}, which link compression is able to alleviate.

Finally, \myName{} provides high \shepherd{performance benefits} for all three classes of workloads with different locality characteristics thanks to synergistically integrating both \LC{} and \PQ{}: (i) \PQ{} helps hide the (de)compression latencies in \LC{} and migrate fewer pages in order to prioritize critical path cache line \shepherd{movements}, and (ii) \LC{} \shepherd{releases network bandwidth resources and} helps recover the lost spatial locality in pages by moving more pages with the available network bandwidth. \emph{dr} and \emph{rs} show only 1.05$\times$ speedup over \RemoteOnly{} as neither \LC{} nor \PQ{} is able to provide speedups due to the poor \shepherd{application} data compressibility and high spatial locality \shepherd{within pages} (which favors moving pages rather than cache lines). \myName{}'s adaptive approach also provides high \shepherd{performance} benefits across all network configurations: (i) when the \netlats{} are high, cache lines movements are slowed down and the sub-block queue fills \shepherd{up} \revised{faster}, %more quickly, 
thus \myName{} favors moving more pages, which is more effective at high network \netlats{}; and (ii) the approximate bandwidth partitioning approach effectively prioritizes cache line over page movements even when network bandwidth is constrained. %In this case, fewer pages are migrated, while critical path cache lines are still prioritized. 
Therefore, \myName{} significantly outperforms the state-of-the-art \RemoteOnly{} scheme 
by 1.85$\times$, 2.36$\times$, 2.97$\times$ for 1/2, 1/4, and 1/8 \bwf{}, respectively.

Overall, we conclude that \myName{}'s cooperative techniques provide a robust approach to alleviate data movement overheads across various network characteristics and application behavior.

\subsection{Memory Access Costs}
\begin{figure}[H]
    \vspace{2pt}
    \includegraphics[width=0.99\linewidth]{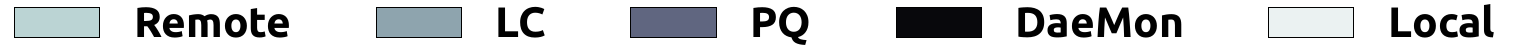}\vspace{7pt}
    \includegraphics[width=\linewidth]{results/singlethread_overview_netlat_100_bwsf_2_tot_latency.pdf}\vspace{6pt}
    \includegraphics[width=\linewidth]{results/singlethread_overview_netlat_400_bwsf_4_tot_latency.pdf}
    \vspace{-12pt}
    \caption{\shepherd{Data} access costs \shepherd{achieved by various schemes} normalized to \RemoteOnly{}.}
    \label{fig:memory-latency}
    \vspace{-2pt}
\end{figure}

%\noindent\textbf{Memory Access Costs.} 
Figure~\ref{fig:memory-latency} compares the average \shepherd{data} access costs (\shepherd{latencies}) \shepherd{achieved by various schemes} normalized to \RemoteOnly. Due to space limitations, in the remaining plots, we \shepherd{present} a \emph{representative} subset of our evaluated workloads, but we report geometric mean values across \emph{all} evaluated workloads. \shepherd{Please also see Figure~\ref{fig:net-bandwidth} in Appendix §~\ref{Appendixbl}, which compares the network bandwidth utilization achieved by the various data movement schemes.}

We make three observations. First, \LC{} improves \shepherd{data} access costs over \RemoteOnly{} by 2.12$\times$ across all network configurations \fixed{(not graphed)}, \shepherd{because} it reduces the network processing costs and \shepherd{queueing} delays by sending fewer bytes through the network. \PQ{} improves access costs (2.06$\times$ over \RemoteOnly{} across all configurations) by prioritizing critical path cache line movements. Second, \PQ{} \emph{significantly} reduces \shepherd{data} access costs in workloads with poor page locality (e.g., \textit{pr, nw}), \shepherd{since} critical path cache line \shepherd{movements} are not stalled by migrating pages. However, in applications with high \shepherd{data} locality (e.g., \textit{dr, rs}), although \PQ{} reduces data access costs by 1.43$\times$ over \RemoteOnly{}, it improves performance by only 1.05$\times$, because \shepherd{the selection granularity unit} favors sending pages for workloads with high locality and \shepherd{a} few requests are served at cache line granularity. Third, \myName{} significantly reduces \shepherd{data} access costs by 3.06$\times$ over \RemoteOnly{}. \myName{} \revised{employs} link compression to \shepherd{migrate} more pages with lower network overhead over \PQ{}, thus exploiting more data locality, while also leveraging the ability to prioritize critical cache line \shepherd{requests}. In \emph{pr}, \myName{} can achieve lower access latency than \LocalOnly{}, since serving requests from both \localMemory{} and \remoteMemory{} increases the effective \shepherd{aggregate} memory bandwidth.

\subsection{Hit Ratio in Local Memory}
\begin{figure}[H]
    %\vspace{-8pt}
    \includegraphics[width=0.76\linewidth]{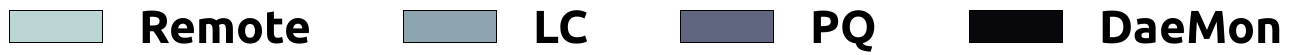}\vspace{6pt}
    \centering
    \includegraphics[width=1\linewidth]{results/singlethread_overview_netlat_100_bwsf_2_hit_rate.pdf}\vspace{6pt}
    \includegraphics[width=1\linewidth]{results/singlethread_overview_netlat_400_bwsf_4_hit_rate.pdf}
    \vspace{-14pt}
    \caption{Hit ratio in \localMemory{} \shepherd{achieved} by various schemes.}
    \label{fig:hit-ratio}
    \vspace{-4pt}
\end{figure}

%\noindent\textbf{Hit Ratio in Local Memory.} 
Figure~\ref{fig:hit-ratio} presents the hit ratio in \localMemory{}, and is thus a measure of the page movement benefits. To prioritize cache lines, \PQ{} throttles some page migrations, thus reducing the \localMemory{} hit rate as a tradeoff for reduced access latencies to critical path cache line \shepherd{requests}. However, \myName{} enables moving more pages \shepherd{over \PQ{} thanks to link compression}, while still retaining the cache line prioritization benefits of \PQ{}. The numbers shown over each bar for \myName{} present the additional pages that were moved in \myName{} as a percentage over \PQ{}, thanks to the reduced bandwidth consumption provided by \revised{link compression}. A zero value indicates that neither \PQ{} nor \myName{} has throttled any page movement.

We \revised{draw three findings.} First, \RemoteOnly{} \revised{has} on average 97.7\% hit ratio in \localMemory. Thanks to high spatial locality, all workloads benefit from page migration, leading to high hit rates: even workloads with relatively poor spatial locality (e.g., \emph{nw}) have 90\% hit ratio in \localMemory. Second, \PQ{} decreases the hit ratio in \localMemory{} by up to 18.4\% over \RemoteOnly{}, because \PQ{} throttles page movements in some workloads to prioritize cache line \shepherd{requests}, thus increasing the \shepherd{number of} accesses to \remoteMemory. Third, \myName{} recovers most of the lost \localMemory{} hits, achieving on average \emph{only} 0.4\% worse hit ratio over \RemoteOnly{}. Leveraging link compression in \myName{} reduces \shepherd{network} bandwidth consumption and significantly increases the number of pages that can be migrated \fixed{over \PQ{}}. %compared to \PQ{}. 
%For example, in \emph{bf, bc} \myName{} recovers the 100\% of \PQ{}'s throttled pages. 
Across all configurations \fixed{(not graphed)} \myName{} migrates 68.9\% of the pages throttled by \PQ{} via leveraging \shepherd{link} compression. %, while retaining the benefits of cache line prioritization. 
We conclude that \myName{} enables both leveraging the benefits of \shepherd{data} locality within pages and the prioritization of critical path cache line \shepherd{requests}.

\subsection{Sensitivity Study to Bandwidth Partitioning Ratio}\label{PqSensitivityBl}
\begin{figure}[t]
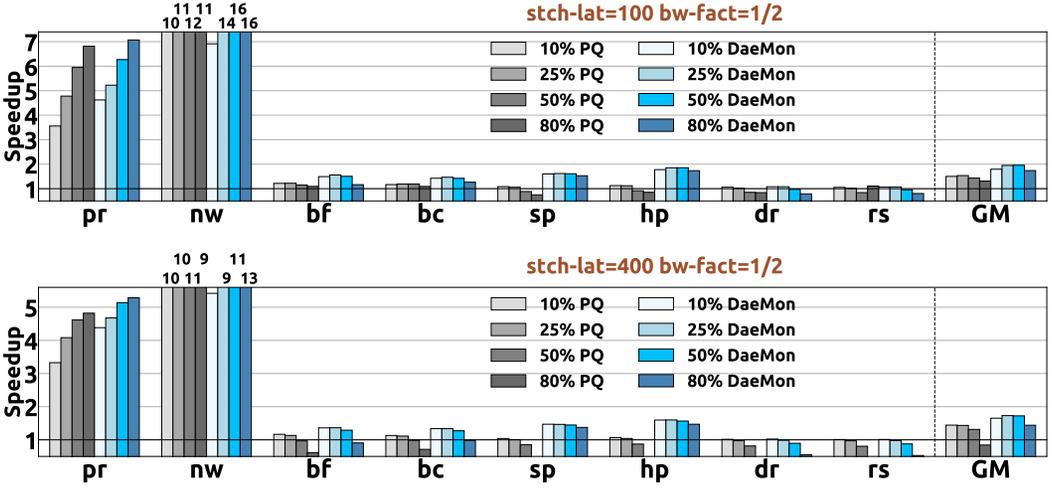

    %\vspace{-8pt}
    \centering
    \includegraphics[width=1\linewidth]{results/pq_sens_netlat_100_bwsf_2_speedup.pdf}\vspace{8pt}
    \includegraphics[width=1\linewidth]{results/pq_sens_netlat_400_bwsf_2_speedup.pdf}
    \vspace{-14pt}
    \caption{Performance of \PQ{} and \myName{} normalized to \RemoteOnly{} varying the bandwidth partitioning ratio.}
    \label{fig:single-pq}
    \vspace{-2pt}
\end{figure}

%\noindent\textbf{Bandwidth Partitioning Ratio.} 
Figure~\ref{fig:single-pq} \shepherd{presents} a sensitivity study on  the bandwidth partitioning ratio between the cache line and page \shepherd{movements}.

We \revised{draw} three \revised{findings}. First, a higher bandwidth partitioning ratio (e.g., 50\%) than \myName{}'s default 25\% ratio, incurs slowdowns in workloads of medium and high spatial locality, and only improves performance in workloads with very low  locality \shepherd{within pages} (e.g., \emph{pr, nw}). This is because high bandwidth partitioning ratios favor cache line movements and throttle a higher number of page movements. Second, since cache line data 
movements are affected more by the \netlat{} compared to page \revised{movements}, the performance benefits of higher bandwidth partitioning ratios reduce at higher \netlats{}. For example, in \emph{pr}, the 50\% bandwidth partitioning ratio outperforms 25\% ratio by 1.19$\times$ and 1.08$\times$ using \myName{} at 100 ns and 400 ns \netlat{}, respectively. Finally, across all different \bwfs{} \fixed{(not graphed)}, \myName{}'s default 25\% ratio outperforms the 50\% ratio by \revised{1.02$\times$} and \revised{1.04$\times$}  for 100 ns and 400 ns \netlat{}, respectively, and the 80\% ratio by \revised{1.07$\times$}  and \revised{1.33$\times$}  for 100 ns and 400 ns \netlat{}, respectively. We conclude that \myName{}'s default 25\% \revised{ratio} on average performs best across all various network and application characteristics.

\subsection{Sensitivity Study to Various Compression Algorithms}\label{CompressionSweepbl}
\begin{figure}[t]
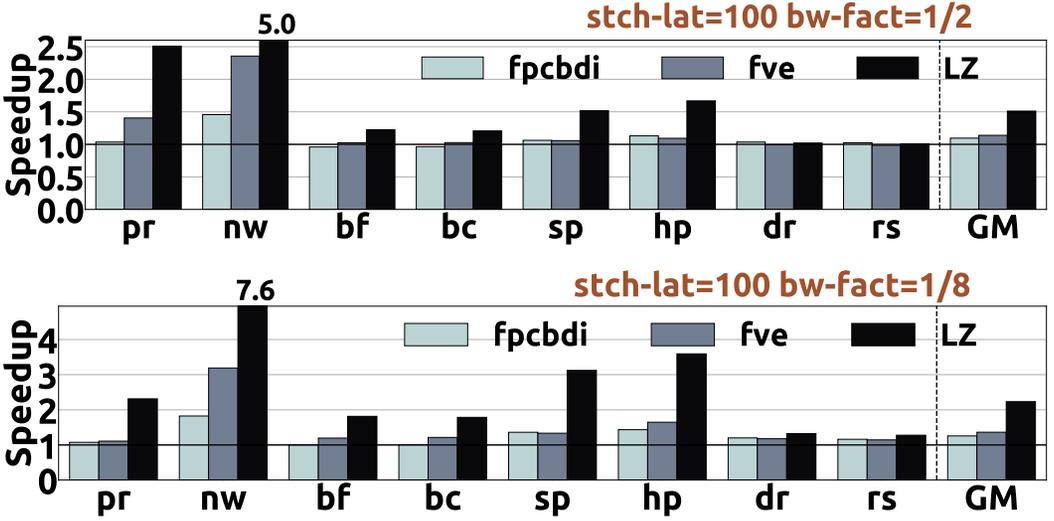

    %\vspace{-6pt}
    \centering
    \includegraphics[width=1\linewidth]{results/compression_sweep_netlat_100_bwsf_2_speedup.pdf}\vspace{8pt}
    \includegraphics[width=1\linewidth]{results/compression_sweep_netlat_100_bwsf_8_speedup.pdf}
    \vspace{-14pt}
    \caption{Performance of \LC{} varying the compression scheme.}
    \label{fig:single-compression}
    \vspace{-3pt}
\end{figure}

%\noindent\textbf{Compression Scheme.} 
Figure~\ref{fig:single-compression} compares the performance of \LC{} normalized to \RemoteOnly{} with three compression schemes: (i) \textit{fpcbdi}: a latency-optimized hybrid scheme of BDI~\cite{Pekhimenko2012BDI} and FPC~\cite{Alameldeen2004FPC} with 4-cycle (de)compression latency per cache line~\cite{Kim2017Transparent}; (ii) \textit{fve}: the latency-optimized FVE ~\cite{Thuresson2008Accommodation} \shepherd{scheme} using a 256B dictionary table and \shepherd{having} 6-cycle (de)compression latency per cache line~\cite{Thuresson2008Accommodation}; and (iii) \textit{LZ}: \myName{}'s compression ratio-optimized LZ-based scheme~\cite{Kim2017Transparent,Tremaine2001MXT} (See details on §~\ref{CompressionBl}).

%We \revised{draw} two \revised{findings}. First, 
We \shepherd{observe that} \textit{LZ} always outperforms \RemoteOnly{}, despite the high (de)compression latencies, \shepherd{because} the network overheads are significantly higher, indicating that link compression is a highly effective solution for \disaggrSystems{}. \textit{dr} and \textit{rs} show little \shepherd{performance} improvement with \textit{LZ}, \shepherd{because} the \shepherd{application} data is less compressible (\shepherd{their compression ratio is} 1.42$\times$ versus 4.47$\times$ on average \shepherd{across all evaluated workloads}). Moreover, \textit{LZ} outperforms \textit{fpcbdi} and \textit{fve} across all network configurations \fixed{(not graphed)} by 1.54$\times$ and 1.44$\times$ on average, 
respectively, \shepherd{since} it \revised{achieves} higher compression ratios (on average 2.92$\times$ and 2.73$\times$ higher \shepherd{compression ratio} than \textit{fpcbdi} and \textit{fve} respectively). The benefits of \textit{LZ} over \textit{fpcbdi} and \textit{fve} are even higher in the more bandwidth limited configurations (e.g., with 1/8 \bwf{}). Therefore, \shepherd{we conclude that} the high network overheads in \disaggrSystems{} favor compression algorithms that provide higher compression ratios, \shepherd{since} the benefits of the reduced bandwidth consumption outweigh the higher (de)compression latencies.

\subsection{Network Disturbance Study}
\begin{figure}[t]
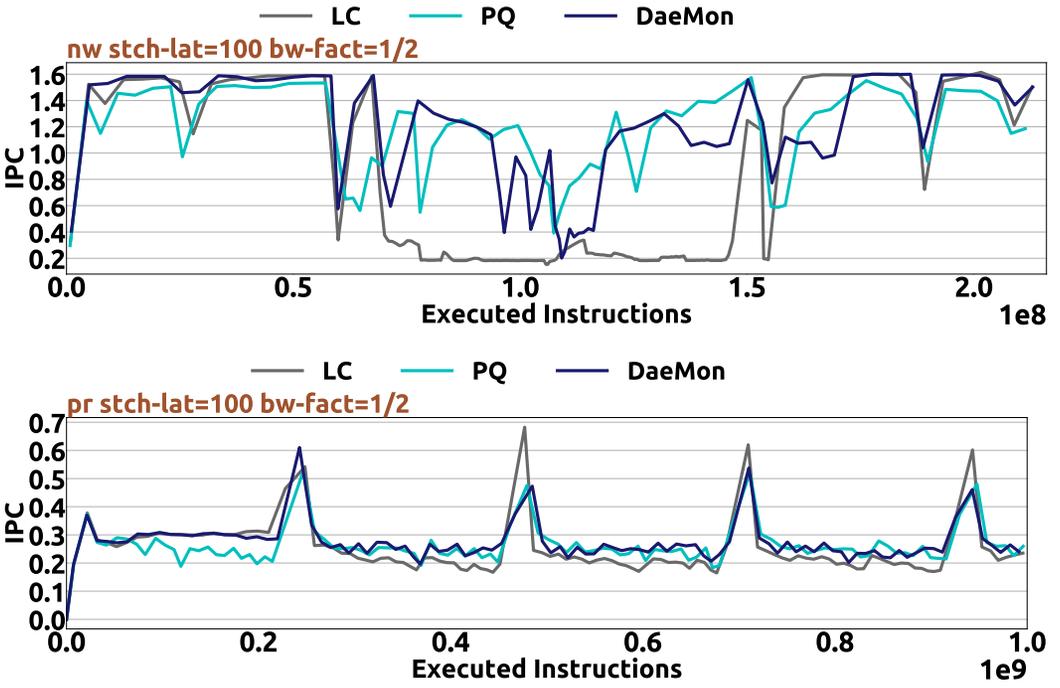

    %\vspace{-6pt}
    \centering
    \includegraphics[width=\linewidth]{results/nw_factor_2_netlat_100_page_size_4096.pdf}\vspace{6pt}
    \includegraphics[width=\linewidth]{results/pr_factor_2_netlat_100_page_size_4096.pdf}
    \vspace{-18pt}
    \caption{Performance of \LC{}, \PQ{}, \myName{}, when creating artificial disturbance in the network during runtime.}
    \label{fig:disturbance}
    \vspace{2pt}
\end{figure}

\begin{figure}[!htb]
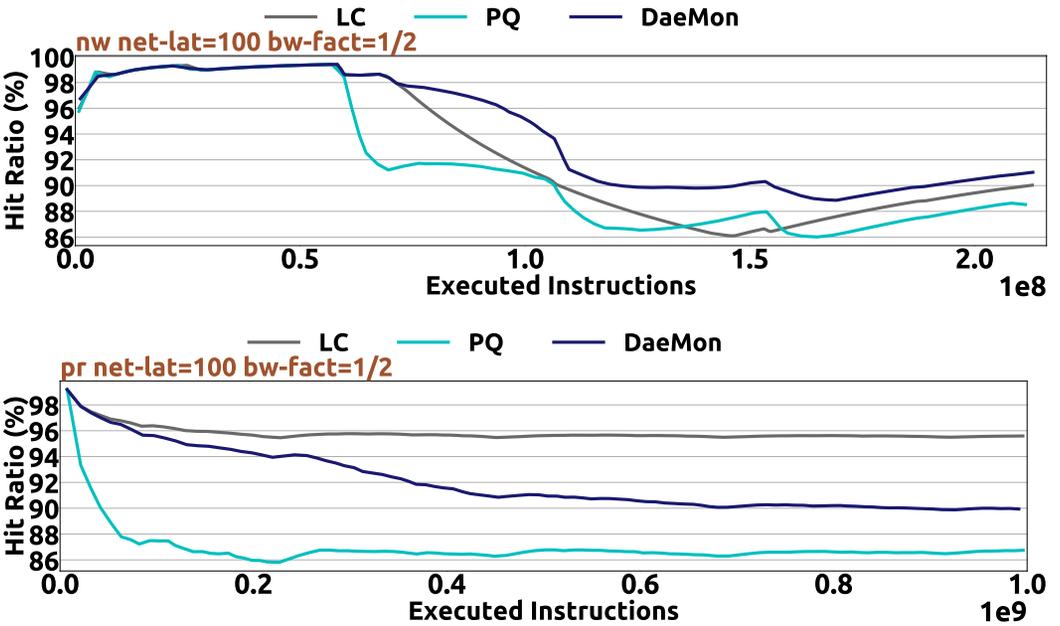

    %\vspace{-6pt}
    \centering
    \includegraphics[width=\linewidth]{results/nw_factor_2_netlat_100_page_size_4096_hitrate.pdf}\vspace{6pt}
    \includegraphics[width=\linewidth]{results/pr_factor_2_netlat_100_page_size_4096_hitrate.pdf} 
    \vspace{-18pt}
    \caption{Hit ratio in \localMemory{} of \LC{}, \PQ{} and \myName{}, when creating artificial disturbance in the network during runtime.}
    \label{fig:disturbance-hit}
    %\vspace{-4pt}
\end{figure}

%\noindent\textbf{Network Disturbance.} 
%Fig.~\ref{fig:disturbance} \revised{shows} the IPC of \LC{}, \PQ{} and \myName{} when the network traffic varies during runtime: we simulate contention from other \computeComponents{} \revised{sharing} the same network, by artificially injecting packets inside the network. We evaluate \emph{pr} and \emph{nw}, as they incur the highest data movement costs. 
Figures~\ref{fig:disturbance} and ~\ref{fig:disturbance-hit} compare the IPC and the hit ratio in \localMemory{} \revised{respectively, } of \LC{}, \PQ{} and \myName{}, when the network traffic varies during runtime: we simulate contention from other \computeComponents{} that share the same network, by artificially injecting packets inside the network. We evaluate \emph{pr} and \emph{nw}, as they incur the highest data movement costs.

\myName{} outperforms both \LC{} and \PQ{} by 2.85$\times$ and 1.19$\times$, respectively, even when network traffic varies during runtime. \myName{}  effectively adapts to varying application behavior and network conditions at runtime. For example, in \emph{nw}, in the first 50M instructions, \myName{} benefits more from \LC{} as the application has high bandwidth consumption and higher locality \shepherd{within pages}. In the next 100M instructions, the workload exhibits less data locality \shepherd{within pages}, and \myName{} benefits \shepherd{more} from \PQ{}, which provides significant performance benefits over \LC{} by effectively prioritizing critical path cache line \shepherd{requests}. In the last part of execution, \myName{} again leverages the benefits of \LC{}. Therefore, \shepherd{we conclude that} \myName{} provides a versatile approach to \revised{dynamic} \shepherd{and variable} runtime application and network characteristics.

\subsection{Multithreaded Performance}
\begin{figure}[t]
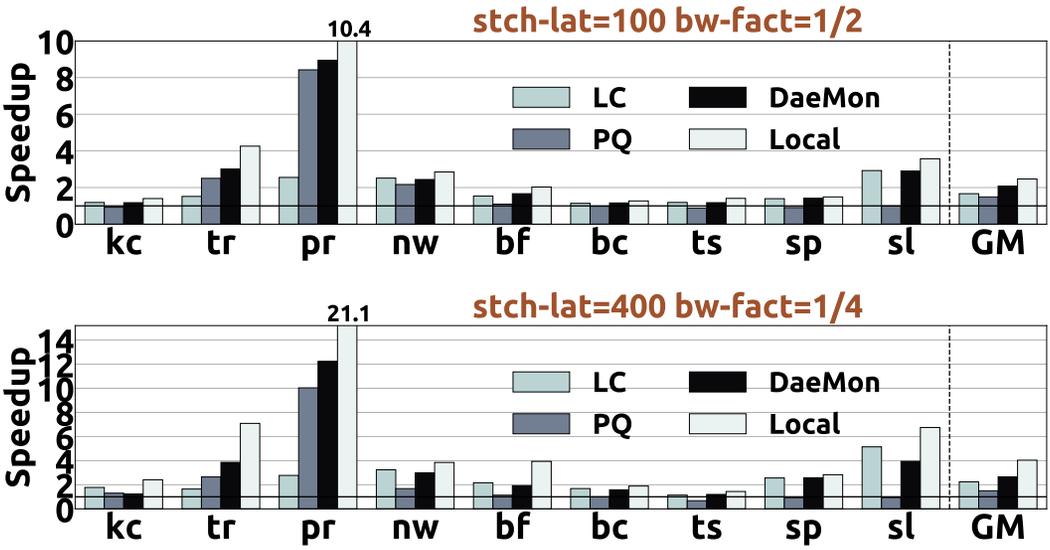

    %\vspace{2pt}
    \centering
    \includegraphics[width=\linewidth]{results/multithread_cores8_netlat_100_bwsf_2_speedup.pdf}\vspace{10pt}
    \includegraphics[width=\linewidth]{results/multithread_cores8_netlat_400_bwsf_4_speedup.pdf}
    \vspace{-14pt}
    \caption{Speedup achieved by various schemes in multithreaded \revised{workloads normalized to} \RemoteOnly{}.}
    \label{fig:multi-speedup}
    \vspace{4pt}
\end{figure}

%\noindent\textbf{Multithreaded Performance.} 
Figure~\ref{fig:multi-speedup} shows \myName{}'s performance benefits for 
multithreaded workloads on 8 OoO cores, \revised{thus evaluating more bandwidth-limited executions compared to that of Figure~\ref{fig:single-performance}.} Please also see Figure~\ref{fig:bandwidth-sens} in \shepherd{Appendix §~\ref{Appendixbl} which evaluates even more bandwidth-limited executions.} 
%Note that the missing workloads in this figure are only supported as singlethreaded. 
Across \emph{all} workloads and network configurations \fixed{(not graphed)}, \myName{} outperforms the typically-used \RemoteOnly{} scheme 
by 2.73$\times$ \shepherd{on average}. \fixed{When network bandwidth is very limited, e.g., 1/16 \bwf{} (Figure~\ref{fig:bandwidth-sens}), \myName{}'s benefits are even higher, by 3.95$\times$ over \RemoteOnly{}.}

\subsection{Sensitivity Study to Replacement Policy in Local Memory}
%\noindent\textbf{FIFO Replacement Policy in Local Memory.} 
Figure~\ref{fig:speedup-fifo} \rebuttal{compares \myName{} and \LocalOnly{} \shepherd{normalized to} \RemoteOnly{}, when using First-In-First-Out (FIFO) replacement policy in \localMemory{}.

\begin{figure}[H]
    %\vspace{-2pt}
    \includegraphics[width=1\linewidth]{results/singlethread_overview_netlat_100_bwsf_2_speedup_fifo.pdf}\vspace{9pt}
    \includegraphics[width=1\linewidth]{results/singlethread_overview_netlat_100_bwsf_4_speedup_fifo.pdf}
    \vspace{-14pt}
    \caption{Performance of \LocalOnly{} and \myName{} over \RemoteOnly{}, when using FIFO replacement policy in \localMemory{}.}
    \label{fig:speedup-fifo}
    %\vspace{-4pt}
\end{figure}

Across \emph{all} workloads and network configurations (not graphed), \myName{} outperforms the widely-adopted \RemoteOnly{} scheme by 2.63$\times$, when using a FIFO replacement policy in \localMemory{}. \myName{} is orthogonal to the replacement policy used in \localMemory, \shepherd{and can be used synergistically with any arbitrary replacement policy in \localMemory{} to even further reduce data access costs}. Overall, \myName{} can significantly mitigate the data movement overheads in fully \disaggrSystems{} independently on the number of data migrations happens during runtime: even when a small number of data migrations happens during runtime (e.g., thanks to sophisticated approaches such as intelligent replacement policies in \localMemory{}, hot page placement/selection techniques, page prefetchers), \myName{} can even further alleviate the data movement costs by dynamically selecting the granularity of data \shepherd{movements}, prioritizing the critical cache line requests, and opportunistically moving compressed pages at slower rates. Therefore, we conclude that \myName{} can work synergistically with sophisticated replacement policies in \localMemory{}, page prefetchers and intelligent page placement/movement techniques to even further improve system performance.}

\subsection{Sensitivity Study to Multiple Memory Components}
%\noindent\textbf{Multiple Memory Components.} 
Figure~\ref{fig:multimem} compares \RemoteOnly{} and \myName{} normalized to 
\LocalOnly{}, \shepherd{when} varying the number of \memoryComponents{} \shepherd{and} having \shepherd{a} different network configuration for each \memoryComponent{}. \rebuttal{Please also see Figure~\ref{fig:multimem-scalability} in Appendix §~\ref{Appendixbl}.} We evaluated distributing memory pages with either a round-robin way or randomly across remote 
\memoryComponents{}, and draw the same key observation for both distributions. When adding more \memoryComponents{} using the same network configuration with that of when having one \memoryComponent{} (e.g., having 100 ns \netlat{} and 1/4 \bwf{} for each \memoryComponent{}), performance of both \RemoteOnly{} and \myName{} improves over \LocalOnly{}: \shepherd{memory} pages are distributed across \textit{multiple} \memoryComponents{} and the system provides larger aggregate network and memory bandwidth, thus data migrations incur smaller overheads.
\shepherd{Finally}, \myName{} significantly outperforms \RemoteOnly{} by 3.25$\times$ across all workload-architecture combinations, and constitutes a \revised{scalable} solution for large-scale \disaggrSystems{} with multiple hardware components and various architectures.

\begin{comment}
\begin{figure}[H]
   \vspace{-11pt}
   \begin{minipage}{\columnwidth}
  \includegraphics[width=1.0\linewidth]{results/multimem.pdf}
  \end{minipage}\hspace{3pt}
  \begin{minipage}{1\columnwidth} \vspace{4pt}
  \hspace{1pt}
  \resizebox{\columnwidth}{!}{%
    \begin{tabular}{c c c c c c c c c c} 
    \toprule
    \textbf{} & \textbf{MC1.1} &   \textbf{MC2.1}  &  \textbf{MC2.2} &  \textbf{MC2.3} &  \textbf{MC4.1} &  \textbf{MC4.2} &  \textbf{MC4.3} &  \textbf{MC4.4}\\ 
    \midrule
    \textbf{\#\memoryComponents{}} & 1 & 2 & 2 & 2 & 4 & 4 & 4 & 4 \\
    \textbf{stch-lat} & 100 & 100-100 &  400-400 & 100-100 & 100-100-100-100 & 100-400-100-400 & 400-400-400-400 & 100-100-100-100 \\
    \textbf{bw-fact} & 1/4 & 1/4-1/4 & 1/4-1/8 & 1/8-1/8 & 1/4-1/4-1/4-1/4 &  1/4-1/8-1/4-1/8  & 1/8-1/8-1/8-1/8 & 1/8-1/16-1/8-1/16 \\
    \bottomrule
    \end{tabular}}
  \end{minipage}%
  \vspace{-8pt}
  \caption{%Performance 
  \revised{Performance} of \RemoteOnly{} and \myName{} over \LocalOnly{} when using multiple \memoryComponents{}. %normalized to \LocalOnly{}.  
  }
   \label{fig:multimem}
  \vspace{-11pt}
\end{figure}
\end{comment}

\begin{figure}[t]
   %\vspace{-2pt}
   \begin{minipage}{\columnwidth}
  \includegraphics[width=1.0\linewidth]{results/multimem.pdf}
  \end{minipage}\vspace{8pt}\hspace{3pt}
  \begin{minipage}{1\columnwidth} \vspace{4pt}
  \centering
  \resizebox{0.88\columnwidth}{!}{%
  \begin{tabular}{c c c c} 
    \toprule
    \textbf{} &  \textbf{\#\memoryComponents{}} & \textbf{stch-lat} & 
    \textbf{bw-fact} \\
    \midrule
    \textbf{MC1.1} & 1 & 100 & 1/4 \\
    \textbf{MC2.1}  & 2 & 100-100 & 1/4-1/4 \\
    \textbf{MC2.2} & 2 & 400-400 & 1/4-1/8 \\
    \textbf{MC2.3} & 2 & 100-100 & 1/8-1/8 \\
    \textbf{MC4.1} & 4 & 100-100-100-100 & 1/4-1/4-1/4-1/4 \\
    \textbf{MC4.2} & 4 & 100-400-100-400 &  1/4-1/8-1/4-1/8 \\
    \textbf{MC4.3} & 4 & 400-400-400-400 & 1/8-1/8-1/8-1/8 \\
    \textbf{MC4.4} & 4 & 100-100-100-100 & 1/8-1/16-1/8-1/16 \\
    \bottomrule
    \end{tabular}}
  \end{minipage}%
  \vspace{-4pt}
  \caption{%Performance 
  \revised{Performance} of \RemoteOnly{} and \myName{} over \LocalOnly{} when using multiple \memoryComponents{}. %normalized to \LocalOnly{}.  
  }
   \label{fig:multimem}
  %\vspace{-6pt}
\end{figure}

\subsection{Sensitivity Study to Multiple \rebuttal{Concurrent} Workloads}
%\noindent\textbf{Multiple \rebuttal{Concurrent} Workloads.}
Figure~\ref{fig:multibench} shows \myName{}'s performance benefits when \rebuttal{concurrently} running multiple %workloads on 4 OoO cores. 
\revised{workloads on a \computeComponent{} with 4 OoO cores.} The performance of \emph{each} core is normalized to that of the same core using \RemoteOnly{}. The \localMemory{} %can fit
\shepherd{hosts} $\sim$15\% and $\sim$9\% of \emph{each} application's working set, when running 2 and 4 workloads, respectively. \myName{} outperforms \RemoteOnly{} by 1.96$\times$ across all multiple-workload experiments, thus being highly efficient and \shepherd{performant} when multiple \emph{heterogeneous} jobs concurrently run in the \disaggrSystem{}.

\begin{figure}[H]
  %\vspace{2pt}
  \includegraphics[width=\linewidth]{results/multibench_speedup_netlat_100_bwsf_2.pdf}
  \vspace{-14pt}
  \caption{Performance of \myName{} over \RemoteOnly{} when running multiple \shepherd{concurrent} workloads in a 4-CPU \computeComponent{} and a \memoryComponent{}.}
   \label{fig:multibench}
  \vspace{-2pt}
\end{figure}

\subsection{Key Takeaways and Recommendations}
This \shepherd{section summarizes our key takeaways and recommendations extracted from our evaluations.}

\noindent \textbf{Key Takeaway \#1.} \shepherd{\textit{There is no one-size-fits-all granularity in data movements: the best-performing granularity at each time depends on the network/system load and the application data access patterns, which can significantly vary across applications and within application during runtime.} Figure~\ref{fig:single-performance} demonstrates that some applications significantly benefit from the prioritization of critical cache line data movements (e.g., \emph{pr, nw}), and some applications only benefit from page migrations that leverage data locality (e.g., \emph{dr, rs}).  Figure~\ref{fig:single-compression} shows that some applications have highly compressible data, and thus greatly benefit from \emph{compressed} page granularity data movements. Finally, Figure~\ref{fig:disturbance} proves that the application behavior and network traffic can highly vary during runtime, and thus the best-performing data movement granularity needs to adapt to the application characteristics and network/system conditions. Therefore, we recommend that system and hardware designers of \disaggrSystems{} implement system-level solutions and hardware mechanisms that dynamically change and adapt their configurations and selection methods to the availability of the system resources and the runtime behavior of the heterogeneous applications. }

\noindent \textbf{Key Takeaway \#2.} \shepherd{\textit{Typical datacenter applications exhibit high data locality within memory pages (e.g., 4KB).} Figure~\ref{fig:hit-ratio} shows that \RemoteOnly{} achieves high data locality, i.e., always has at least 90\% hit ratio in \localMemory{},  across a wide variety of datacenter workloads with diverse access patterns. Therefore, migrating data at a large granularity, e.g., page granularity, is very effective and critical to achieving high system performance in fully \disaggrSystems{}. To this end, we suggest that hardware and system designers of \disaggrSystems{} retain coarse-grained data migration (i.e., page granularity data migration), since it both enables high performance and maintains low metadata overheads for address translation in \localMemory{} and \remoteMemory.}

\noindent \textbf{Key Takeaway \#3.} \shepherd{\textit{Aggressively prioritizing the cache line granularity data movements that are on the critical path might hurt performance.} Figure~\ref{fig:single-pq} shows that a high bandwidth partitioning ratio, e.g., 50\% or 80\% bandwidth partitioning ratio, which significantly prioritizes the cache line granularity data movements over the page granularity data movements, incurs significant performance slowdowns in workloads with medium and high spatial locality. %, since effective page movements are aggressively stalled. 
As a result, we suggest that hardware and system designers of data movement solutions tailored for \disaggrSystems{} always ensure that page migrations are not aggressively stalled.}

\noindent \textbf{Key Takeaway \#4.} \shepherd{\textit{Distributed and disaggregated data movements solutions are highly effective and efficient in fully \disaggrSystems{}.} \disaggrSystems{} are distributed architectures and comprise multiple hardware devices, each of them is independently and transparently managed from other hardware components in the system. Our evaluations in Figures~\ref{fig:multimem} and ~\ref{fig:multimem-scalability} show that  distributed and disaggregated solutions for data movement (i.e., \myName{}) better leverage the available aggregate network and memory bandwidth in the system, and enable high scalability to large-scale \disaggrSystems{} with multiple hardware components. To this end, we recommend that hardware architects design distributed hardware mechanisms for fully \disaggrSystems{}.}

%\vspace{-5pt}
\section{Related Work}\label{RelatedWorkbl}
%\vspace{-3pt}

To our knowledge, this is first work to (i) analyze and alleviate the data movement problem in fully \disaggrSystems; (ii) enable prioritized and decoupled movement of data at multiple granularities simultaneously to reduce access latencies; (iii) propose a dynamic selection granularity mechanism with approximate bandwidth partitioning to effectively leverage both cache line and page movement depending on application and network characteristics; and (iv) implement a synergistic solution of link compression, bandwidth partitioning, and adaptive granularity selection in data movements. We discuss prior work.

\noindent\textbf{Disaggregated Systems.} Several 
prior works~\cite{Shan2018LegoOS,Angel2020Disaggregation,Nitu2018Zombieland,Han2013Network,Calciu2021Rethinking,Aguilera2017Remote,Zhang2020RethinkingDM,Lim2012System,Chenxi2020Semeru,Pengfei2021OneSided,Bindschaedler2020Hailstorm,Peng2020Underutilization,Gu2017Infiniswap,Aguilera2018Remote, Lee2021MIND,Pinto2020ThymesisFlow,Katrinis2016dredbox,guo2021clio,Buragohain2017DiME,Gao2016Network,Rao2016IsMemory,Zervas2018Optically,Gouk2022Direct,Zhou2022Carbink} propose OS modules, system-level solutions, programming frameworks, software management systems, architectures and emulators for \disaggrSystems{}. These works do not tackle \shepherd{the} data movement \shepherd{problem} in \disaggrSystems{}, and thus \myName{} is orthogonal to these \shepherd{proposals}.
MIND~\cite{Lee2021MIND} proposes memory sharing among \computeComponents{} by implementing coherence and address translation in network switches. Kona~\cite{Calciu2021Rethinking} is a software runtime to track cache line granularity accesses to \remoteMemory{}, and eliminate page faults by decoupling the application memory access tracking from the virtual memory page size. However, 
Kona and MIND do not mitigate data movement overheads in \disaggrSystems{}, as data is \shepherd{always} moved at page granularity. Thus, \myName{} is largely orthogonal to these works and could be used to further improve performance. Clio~\cite{guo2021clio} proposes a \disaggrSystem{} that virtualizes and manages \remoteMemory{} at the hardware level (independently to \computeComponents{}), and eliminates expensive page faults in \memoryComponents{}. Clio accesses \emph{remote} data at a byte granularity via dedicated API, however \emph{not} being transparent to programmers. As explained in §~\ref{MotivationPlotbl}, moving data \emph{always} at a small granularity can cause significant performance \shepherd{penalties} in many applications, and does not provide robustness against fluctuations in network characteristics. Instead, \myName{} is software-transparent, robust and significantly alleviates data movement costs via decoupled and selective data movement at \emph{multiple} granularities. Lim et al.~\cite{Lim2009Disaggregated} propose a disaggregated architecture and characterize moving data \emph{only} at %a 
cache line or page granularity. The authors show that the page-based configuration outperforms the cache line configuration at most common patterns (as observed in §~\ref{MotivationPlotbl}), however it does not address the high performance penalties of page migrations. \rebuttal{Maruf and Chowdhury~\cite{Maruf2020Effectively} propose a page prefetching scheme for \disaggrSystems{}, which however can only help applications with high locality within pages, and does not capture the significant variability in data access costs of fully \disaggrSystems{}. \myName{} is orthogonal to page prefetchers and can work synergistically with them to even further improve performance, as described in Section~\ref{ExtensionsBl}. We leave the experimentation of their synergy for future work.  }

\noindent\textbf{Hybrid Memory Systems.}
Numerous works for hybrid memory 
systems propose %efficient 
data placement schemes~\cite{Agarwal2015Page,Chou2017Batman,Dulloor2016Data,Ruan2020AIFM,Kim2021Exploring,Doudali2021Cori,Lei2019Hierarchical,Sudarsun2017HeteroOS,TMO2022Weiner,Chang2021Dynamap,Feeley1995Workstation}, or selection methods~\cite{Dong2010Simple,Doudali2019Kleio,Yan2019Nimble,Agarwal2017Thermostat,Mitesh2015Heterogeneous,Liu2017Hardware,Kotra2018Chameleon,Jiang2010CHOP,Vasilakis2019LLCGuided,Prodromou2017MemPod,Singh2022Sibyl,Lagar2019Software} to identify hot memory pages that are migrated to die-stacked DRAM, %which 
that is organized as a cache of a larger main memory. % tier. 
Compared to these approaches, first, intelligent page placement/movement is orthogonal to \myName{}, and cannot by itself address the high overheads caused by remote page migrations across the network, that can be significantly slower than that within the server and more latency/bandwidth-constrained in the context of \shepherd{fully} \disaggrSystems{}.
Second, these prior works 
assume a monolithic centralized system where TLBs/page tables can be leveraged to track page hotness of remote pages (e.g.,~\cite{Dong2010Simple,Doudali2019Kleio,Yan2019Nimble,Liu2017Hardware,Kotra2018Chameleon,Agarwal2017Thermostat,Mitesh2015Heterogeneous}) or that memory allocation/placement is handled by the server itself (e.g.,~\cite{Ruan2020AIFM,Dulloor2016Data,Agarwal2015Page,Kim2021Exploring,Doudali2021Cori,Lei2019Hierarchical,Sudarsun2017HeteroOS,Li2017Utility,Feeley1995Workstation}). However, in \disaggrSystems{}, address translation and memory management are distributed across \memoryComponents{} and cannot be used to track pages at the CPU server side, while \computeComponents{} and \memoryComponents{} are managed by independent kernel monitors that have no visibility/control of other components or data management/placement across components. Similarly, hardware-based approaches ~\cite{Vasilakis2019LLCGuided,Jiang2010CHOP,Prodromou2017MemPod,Li2017Utility} for hybrid systems add \emph{centralized} hardware units at the server side to store page tracking metadata for the second-tier main memory. For example, Chop~\cite{Jiang2010CHOP} adds 4MB of metadata to track 16GB of second-tier memory. These schemes would incur significant area overheads (in the order of GBs) to track large amounts of \remoteMemory{} (in the order of TBs) enabled by \disaggrSystems{}~\cite{Shan2018LegoOS}. Requiring \emph{each} \computeComponent{} to  track a large number of pages enabled by multiple remote \memoryComponents{} would cause scalability issues
%impose significant scalability challenges 
and \shepherd{significantly} limit the benefits of resource disaggregation. \fixed{Thus, designing an effective scalable hot page selection scheme for fully \disaggrSystems{} is an open challenge, and \myName{} could work in conjunction with such schemes to further improve performance.}
%Similarly, hardware-based approaches \revised{introduce centralized hardware units at the server side to store} page tracking metadata at the compute unit (e.g.,~\cite{Vasilakis2019LLCGuided,Jiang2010CHOP,Prodromou2017MemPod,Li2017Utility}), \revised{and would also} incur significant \revised{area overheads (in the orders of GBs)} to track large working sets \revised{(in the order of TBs)} enabled by \disaggrSystems{}. 
Third, all these prior works do not handle variability in data access costs of \disaggrSystems{}. \disaggrSystems{} necessitate an adaptive mechanism given the significant variations in access latencies and bandwidth. 
%based on the network architecture and the location of the remote \memoryComponents{}. %memory module(s). 
Fourth, %note that 
applying/adapting the design of prior schemes tailored 
for tightly-integrated hybrid systems in \disaggrSystems{} might incur significantly higher overheads and require important modifications than that  described in the original papers. 
%Finally, accesses across the network can be significantly slower than that within the server, thus more intelligent page placement/movement cannot by itself address these \revised{high} overheads, necessitating cache line prioritization. 
%\cgiannou{Finally, designing a scalable hot page selection scheme for Disaggregated systems is open challenge:  requiring each CC, to manage/track a large number of pages enabled by multiple remote MCs would impose scalability challenges and significantly limit the benefits of resource disaggregation, while our approach DaeMon could  work in conjunction with such approach to further improve performance via  data compression, prioritization of critical data, and adaptive granularity data movements.}

A few recent works design hardware schemes for commodity servers to enable moving data \emph{only} at cache line granularity
~\cite{Loh2011Efficiently,Loh2012Supporting,Chou2014Cameo,Vasilakis2020Hybrid2} or a larger sub-block granularity (a few cache lines)~\cite{Ryoo2017Silcfm,Jevdjic2013Footprint}. Ekman et al.~\cite{Ekman2005ACost} evaluate a critical-block first approach, 
where each 8KB page is split in blocks of 2KB data, and the requested (critical) 2KB block of data 
is transferred first, and written in DRAM cache. As we show in §~\ref{MotivationPlotbl}, moving data at a single granularity (page or cache line) can incur high performance costs and does not provide robustness towards significant variations in network bandwidth and latencies. %Lagar-Cavilla et al.~\cite{Lagar2019Software} \revised{split the server's main memory in two partitions, i.e., near and far memory, and } identify  cold pages \revised{using page tables} that are transparently compressed and \revised{stored in} far memory \revised{partition}. However, this architecture is not fully disaggregated and \revised{the application's footprint} is expected to fit in the server's \revised{main memory}, which is not the case with \disaggrSystems.

\noindent\textbf{Hardware Compression.} Prior works  propose compression schemes~\cite{Pekhimenko2012BDI,Kim2016BPC,Alameldeen2004FPC,Thuresson2008Accommodation,Chen2010CPack,Yang2000Frequent,Yang2004Frequent,Arelakis2014SC2,Nguyen2015MORC,Nguyen2018Cable,Shafiee2014MemZip,Young2019Enabling,Ekman2005RMC,Pekhimenko2013LCP,Choukse2018Compresso,Qian2018CMH,Wilson1999TheCF,Kjelso2000XMatch,Tian2014Last,Park2021BCD,Abali2001MXT} for cache memory, main 
memory and memory 
bus links in CPUs/GPUs~\cite{Mittal2016Survey,Thuresson2008Memory,Sathish2012Lossless,Vijaykumar2015CABA}, and selection methods 
%\revised{for dynamic data compression}~\cite{Alameldeen2004Adaptive,Alameldeen2007Interactions,Tuduce2005Adaptive,Arelakis2015HyComp,Kim2017Transparent}.
to dynamically enable/disable compression~\cite{Alameldeen2004Adaptive,Alameldeen2007Interactions,Tuduce2005Adaptive}, or find the best-performing compression scheme~\cite{Arelakis2015HyComp,Kim2017Transparent}. 
%These works integrate ratio-optimized or latency-optimized compression schemes depending on the particular context and system's characteristics that they target. Our work leverages link compression to reduce the page movement cost across the network synergistically with decoupled data movement at two granularities, which allows us to tolerate the high compression latencies of ratio-optimized algorithms such as LZ~\cite{Ziv1977LZ}.
These works integrate ratio-optimized or latency-optimized compression schemes depending on the particular context and system's characteristics %that 
they target. Our work enables link compression in page \shepherd{movements} synergistically with decoupled multiple granularity data movement, which allows us to tolerate the high compression latencies of ratio-optimized \shepherd{compression} schemes such as LZ~\cite{Ziv1977LZ}.

%\vspace{-4pt}
\section{Conclusion}
%\vspace{-4pt}

\myName{} is the first adaptive data movement solution for \shepherd{fully} \disaggrSystems{}. \myName{} supports low-cost page migration, \revised{scales elastically to \fixed{multiple} hardware components,} enables software transparency, and provides robustness across various architecture/network characteristics and the application behavior by effectively monitoring \emph{pending} cache line and page movements. Our evaluations \shepherd{using a state-of-the-art simulator} show that \myName{} significantly improves system 
performance and data access costs for a wide range of applications under various architecture and network configurations, and when multiple \revised{jobs} are \revised{simultaneously} running in the system. We conclude that \myName{} is an efficient, \revised{scalable} and robust solution to alleviate data movement overheads in \disaggrSystems{}, and hope that this work encourages further studies of the data movement problem in \disaggrSystems{}.

\begin{comment}

\myName{} is the first adaptive data movement solution for \disaggrSystems{}. \myName{} supports low-cost page migration, \revised{scales elastically to \fixed{multiple} hardware components,} enables software transparency, and provides robustness across various architecture/network characteristics and the application behavior. %by effectively monitoring \emph{pending} cache line and page movements. 
Our evaluations show that \myName{} significantly improves system performance and data access costs for a wide range of applications under various architecture, network configurations, and when multiple \revised{jobs} are \revised{simultaneously} running in the system. We 
%conclude that \myName{} is an efficient, \revised{scalable} and robust solution to alleviate data movement overheads in \disaggrSystems{}, and 
hope that this work encourages further studies of the data movement problem in disaggregated data centers.

\end{comment}

\begin{acks}
We thank the anonymous reviewers from SIGMETRICS 2023, and our shepherd, Abhishek Chandra, for their comments and suggestions. We also thank Konstantinos Kanellopoulos and Ivan Fernandez for their help on technical aspects of this work. 
\end{acks}

\newpage

\bibliographystyle{ACM-Reference-Format}
\bibliography{references}

\newpage

\section*{{\Large APPENDIX}}
\appendix

\section{Extended Results}\label{Appendixbl}

\subsection{Network Bandwidth Utilization} 
\begin{figure}[H]
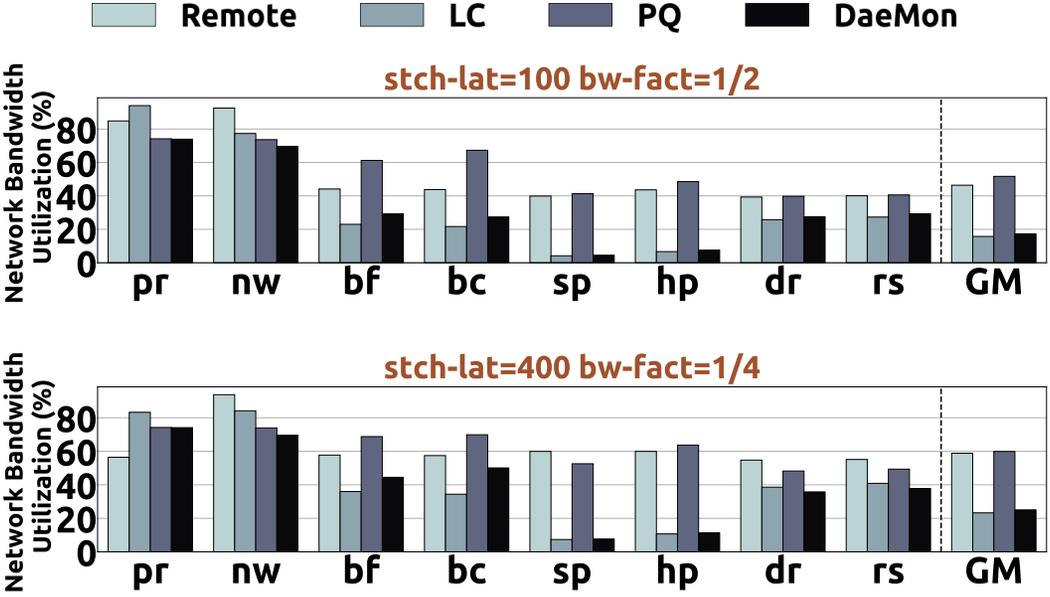

    %\vspace{-2pt}
    \includegraphics[width=0.84\linewidth]{results/legend_hit_rate.png}\vspace{10pt}\hspace{200pt}
    \centering
    \includegraphics[width=1\linewidth]{results/singlethread_overview_netlat_100_bwsf_2_net_bandwidth.pdf}\vspace{18pt}
    \vspace{30pt}\includegraphics[width=1\linewidth]{results/singlethread_overview_netlat_400_bwsf_4_net_bandwidth.pdf}
    \vspace{-34pt}
    \caption{Bandwidth utilization (\%) across the network of a \computeComponent{} and a \memoryComponent{} achieved by various data movement schemes.}
    \label{fig:net-bandwidth}
    \vspace{-2pt}
\end{figure}

Figure~\ref{fig:net-bandwidth} \rebuttal{compares the bandwidth utilization across the network of a \computeComponent{} and a \memoryComponent{} achieved by various data movement schemes.}

We make three \rebuttal{key observations. First, \LC{} typically reduces the network bandwidth utilization over \RemoteOnly{} (by 2.49$\times$ on average across all workloads and network configurations), because fewer bytes are transferred through the network, since remote pages are migrated in a compressed format. Note that \LC{} improves the total execution time over \RemoteOnly{}, and thus in a few workloads, e.g., \emph{pr}, the network bandwidth utilization might be higher within a smaller execution time. Second, \PQ{} decreases the network bandwidth utilization over \RemoteOnly{} in workloads with poor spatial locality within pages (e.g., \emph{nw}), since the selection granularity unit effectively schedules more cache line \shepherd{movements} and fewer page migrations. Instead, \PQ{} might slightly increase the network bandwidth utilization over \RemoteOnly{} in workloads with medium spatial locality within pages (e.g., \emph{bf}, \emph{bc}), since the selection granularity unit enables both cache line and page migrations to leverage both the ability to prioritize critical cache line requests and the benefits of data locality within pages. In workloads with high spatial locality within pages (e.g., \emph{dr}, \emph{rs}), \PQ{} favors more page migrations and fewer cache line \shepherd{movements}, thus achieving similar network bandwidth utilization to \RemoteOnly{}. Third, \myName{} greatly decreases the network bandwidth utilization over \RemoteOnly{} by 2.32$\times$ on average across all workloads and network configurations (not graphed). \myName{} effectively transfers remote pages in a compressed format and on-the-fly selects the granularity of data migrations to \emph{significantly} reduce the bandwidth consumption across the network of fully \disaggrSystems{}. }

%\vspace{-1pt}
\subsection{Sensitivity Study to Switch Latency}

\begin{figure}[H]
    \vspace{-2pt}
    \centering
    \includegraphics[width=\linewidth]{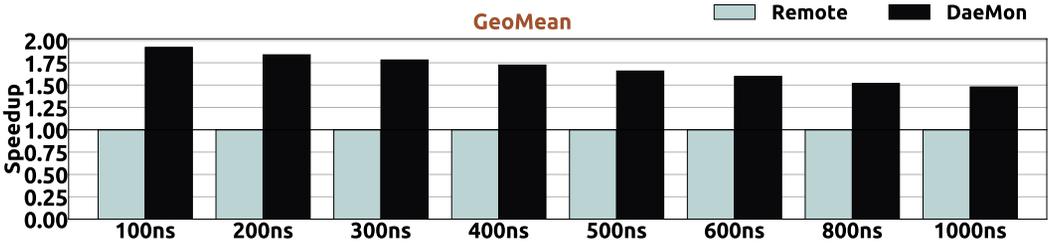}
    \vspace{-14pt}
    \caption{Performance benefits of \myName{} over \RemoteOnly{}, when varying the \netlat{} of the network.}
    \label{fig:netlat-sens}
    %\vspace{-8pt}
\end{figure}

Figure~\ref{fig:netlat-sens} \rebuttal{compares \myName{}'s performance over \RemoteOnly{}'s performance \shepherd{averaged across all workloads,} when varying the \netlat{} of the network. When the fixed \netlat{} becomes very high dominating the total data movement costs, \myName{} has lower benefits over \RemoteOnly{}, since \myName{} does not hide the propagation and switching delays in network components (e.g., fixed processing costs of the packet inside network switches). However, even with a very high \netlat{} in the order of microsecond, i.e., 1$\mu$s (=1000 ns), \myName{} outperforms \RemoteOnly{} by 1.49$\times$ on average  across all workloads.}

\subsection{Sensitivity Study to Network Bandwidth}

To evaluate \rebuttal{bandwidth-limited scenarios, Figure~\ref{fig:bandwidth-sens} compares \myName{}'s performance normalized to \RemoteOnly{}'s performance in  multithreaded workloads running on 8 OoO cores of a \computeComponent{}, when varying the \bwf{} of the network, e.g., up to having a very low bandwidth factor of 1/16 (i.e., network bandwidth is 16$\times$ slower than the DRAM bus bandwidth) \shepherd{between a \computeComponent{} and \memoryComponent{}.} We find that on average \myName{}'s benefits increase over the widely-adopted approach of moving data at page granularity, i.e., \RemoteOnly{}, since \myName{} even more significantly alleviates bandwidth bottlenecks and data movement overheads under bandwidth-constrained scenarios. }

\begin{figure}[H]
    %\vspace{-4pt}
    \centering
    \includegraphics[width=\linewidth]{results/bandwidth_factor_scalability.pdf}
    \vspace{-14pt}
    \caption{Performance benefits of \myName{} normalized to \RemoteOnly{} using multithreaded workloads, when varying the \bwf{} of the network \shepherd{between a \computeComponent{} and \memoryComponent{}.}}
    \label{fig:bandwidth-sens}
    %\vspace{-4pt}
\end{figure}

\subsection{Performance Benefits With Multiple Memory Components}
Figure~\ref{fig:multimem-scalability} \rebuttal{evaluates the performance of \myName{} normalized to \RemoteOnly{}'s performance, when increasing the number of \memoryComponents{} in the system having the same network configuration for each \memoryComponent{}, i.e., 100 ns \netlat{} and a \bwf{} of 1/4. We evaluated distributing memory pages with either a round-robin way or randomly across multiple remote \memoryComponents{}, and drew the same key observations for both distributions. Similarly to Figure~\ref{fig:multimem}, we observe that when pages are distributed across \textit{multiple} \memoryComponents{} and the system provides larger aggregate network and memory bandwidth, data access costs decrease. For example, when increasing the number of \memoryComponents{} from 2 to 4, the remote data access latency decreases by 1.39$\times$ on average across all workloads. However, even when data access costs affect less the total execution time of applications, \myName{} still further mitigates data access overheads: \myName{} outperforms the widely-adopted \RemoteOnly{} approach by 2.09$\times$ and 1.88$\times$ on average across all workloads, when using 2 and 4 \memoryComponents{}, respectively.}

\begin{figure}[H]
    %\vspace{-4pt}
    \centering
    \includegraphics[width=\linewidth]{results/multimem_scalability.pdf}
    \vspace{-14pt}
    \caption{Performance benefits of \myName{} normalized to \RemoteOnly{}, when increasing the number of \memoryComponents{} having 100 ns \netlat{} and a \bwf{} of 1/4 for each \memoryComponent{}.}
    \label{fig:multimem-scalability}
    %\vspace{-8pt}
\end{figure}

\end{document}